\documentclass[aps, prd]{revtex4-2}
\usepackage{xcolor}
\usepackage{float}
\usepackage{amsmath,amsfonts,amssymb,epsfig,graphicx}
\usepackage{slashed}
\usepackage{hyperref}
\usepackage[normalem]{ulem}
\usepackage{lineno}
\usepackage{comment}
\usepackage{afterpage}
\usepackage{titlesec}

\usepackage{subcaption} 
\usepackage{setspace}
\titleformat{\section}{\normalfont\Large\bfseries}{\thesection}{1em}{}

\newcommand{\eg}{{\it e.g.}}

\newcommand{\GeV}{\,\text{GeV}}

\newcommand{\deltaP}{\Delta P_{\alpha\beta}^{\rm CP}}
\newcommand{\pmutau}{P(\nu_{\mu}\to\nu_{\tau})}
\newcommand{\pmuel}{P(\nu_{\mu}\rightarrow\nu_{e})}
\newcommand{\pelmu}{P(\nu_{e}\rightarrow\nu_{\mu})}
\newcommand{\peltau}{P(\nu_{e}\rightarrow\nu_{\tau})}

\newcommand{\lb}{\left[}
\newcommand{\rb}{\right]}
\newcommand{\lp}{\left(}
\newcommand{\rp}{\right)}

\newcommand{\dcp}{\delta_{\rm CP}}

\newcommand{\globes}{\texttt{GLoBES}~}
\newcommand{\genie}{\texttt{GENIE}}

\graphicspath{{./pic/}}

\begin{document}
\title{Muon Beam for Neutrino CP Violation:  connecting energy and neutrino frontiers}

\author{Alim \surname{Ruzi}}
\email[]{alim.ruzi@pku.edu.cn}
\affiliation{State Key Laboratory of Nuclear Physics and Technology, School of Physics, Peking University, Beijing, 100871, China}

\author{Tianyi \surname{Yang}}
\email[]{tyyang99@pku.edu.cn}
\affiliation{State Key Laboratory of Nuclear Physics and Technology, School of Physics, Peking University, Beijing, 100871, China}

\author{Dawei \surname{Fu}}
\email[]{fudw@pku.edu.cn}
\affiliation{State Key Laboratory of Nuclear Physics and Technology, School of Physics, Peking University, Beijing, 100871, China}

\author{Sitian \surname{Qian}}
\email[]{stqian@pku.edu.cn }
\affiliation{State Key Laboratory of Nuclear Physics and Technology, School of Physics, Peking University, Beijing, 100871, China}

\author{Leyun \surname{Gao}}
\affiliation{State Key Laboratory of Nuclear Physics and Technology, School of Physics, Peking University, Beijing, 100871, China}

\author{Qiang \surname{Li}}
\email[]{qliphy0@pku.edu.cn}
\affiliation{State Key Laboratory of Nuclear Physics and Technology, School of Physics, Peking University, Beijing, 100871, China}

\begin{abstract}
We propose here a proposal to connect neutrino and energy frontiers, by exploiting collimated muon beams for neutrino oscillations, which generate symmetric neutrino and antineutrino sources: $\mu^+\rightarrow e^+\,\bar{\nu}_{\mu}\, \nu_{e}$ and $\mu^-\rightarrow e^-\, \nu_{\mu} \,\bar{\nu}_{e}$.
Interfacing with long baseline neutrino detectors such as DUNE and T2K, this experiment can be applicable to measure tau neutrino properties, and also to probe neutrino CP phase, by measuring muon electron (anti-)neutrino mixing or tau (anti-)neutrino appearance, and differences between neutrino and antineutrino rates. 
There are several significant benefits leading to large neutrino flux and high sensitivity on CP phase, including 1) collimated and manipulable muon beams, which lead to a larger acceptance of neutrino sources in the far detector side; 2) symmetric $\mu^+$ and $\mu^-$ beams, and thus symmetric neutrino and antineutrino sources, which make this proposal ideally useful for measuring neutrino CP violation.
More importantly, $\bar{\nu}_{e,\mu}\rightarrow\bar{\nu}_\tau$ and $\nu_{e,\mu}\rightarrow \nu_\tau$, and, $\bar{\nu}_{e}\rightarrow\bar{\nu}_\mu$ and $\nu_{e}\rightarrow \nu_\mu$ oscillation signals can be collected simultaneously, with no needs for separate specific runs for neutrinos or antineutrinos. 
Based on a simulation of neutrino oscillation experiment, we estimate $10^4$ tau (anti-) neutrinos can be collected within 5 years which makes this proposal suitable for a brighter tau neutrino factory. Moreover, more than 7 standard deviations of sensitivity can be reached for $\dcp = |\pi/2|$, within only five ears of data taking, by combining tau and muon (anti-) neutrino appearances. With the development of a more intensive muon beam targeting future muon collider, the neutrino potential of the current proposal will surely be further improved.
\end{abstract}

\maketitle

\section{Introduction} \label{sec:intro}

Novel collision methods and rich phenomena are crucial to keeping high-energy collision physics more robust and attractive~\cite{Lu:2022ibc}. Recent years have witnessed vast development towards next generation high energy colliders, including various proposals on Higgs factory~\cite{EuropeanStrategy:2019qin,Bagger:2022zyy}, revived interest in Muon collider~\cite{Aime:2022flm,MuonCollider:2022nsa,MuonCollider:2022glg, MuonCollider:2022xlm,Quigg:1997uk}, etc.

As for the muon collider design, we take positron on target method (LEMMA) as an example, which has been proposed for high quality muon beam production~\cite{Antonelli:2015nla,Alesini:2019tlf} (Earlier studies using proton on target muon beams can be found in Refs.\cite{Tang:2021lyn,Burguet-Castell:2001ppm,Cao:2014bea} )
Although it is still quite challenging to achieve enough high luminosity for muon beam collisions~\cite{MuonCollider:2022glg, MuonCollider:2022xlm}, we find it quite promising for neutrino oscillation studies, with comparable or even larger neutrino flux than other long baseline neutrino experiments, with more details to be discussed below. In the LEMMA approach, the incident positron energy is around 45 GeV, producing collimated muon pairs with opening angles of around 0.005 rad. and a large boost about $\gamma \sim 200$, which extends the muon lifetime by the order of $\mathcal{O}(10^2)$. Generally, the number of muon pairs produced per positron bunch on target can be expressed as
\begin{equation}
    n(\mu^+\mu^-) = n^+\rho_{e^-}l\sigma(\mu^+\mu^-)
    \label{eq:murate}
\end{equation}
where $n^+$ is the number of $e^+$ in each positron bunch, $\rho_{e^-}$ is the electron density in the medium, $l$ is the thickness of the the target, and $\sigma(\mu^+\mu^-)$ being the cross section of the muon pair production. The number of muon pairs per positron bunch on target can be maximally estimated as $n(\mu^+ \mu^-)_\mathrm{max}\approx n^+\times 10^{-5}$.

Neutrinos are among the most abundant and least understood of all particles in the SM that make up our universe. The neutrino physics has made significant progress in the past few decades. Studies on the neutrino physics are full of novel discoveries, one of them is the observation of neutrino oscillations~\cite{Bilenky:1998dt, Nunokawa:2007qh, Blennow:2013rca}, confirming that at least two types of SM neutrinos have a tiny, but strictly nonzero mass. In addition to this, neutrino oscillations also solved the mysterious solar and atmospheric-neutrino problem~\cite{Mikheyev:1985zog, Bethe:1986ej, Haxton:2012wfz, Maltoni:2015kca, Fogli:2003th}.  The neutrino oscillations in the presence of three active neutrino flavors are described by the mass square differences \rm{i.e.}  $\Delta m_{21}^2 = m_2^2 - m_1^2$, $\Delta m_{31}^2 = m_3^2 - m_1^2$, three mixing angles, $\theta_{12}$, $\theta_{23}$, and $\theta_{13}$, and one Dirac phase, $\dcp$. There are another two parameters: the Majorana phases, $\delta_1$, $\delta_2$. They only play a part in the neutrinoless double beta decay~\cite{Vergados:2012xy} and are directly related to the neutrino nature. The mixing angles and the phases are the elements of a unitary matrix called Pontecorvo-Maki-Nakagawa-Sakata (PMNS) matrix~\cite{Maki:1962mu, Pontecorvo:1967fh}. The available experiments on neutrino oscillations to date have measured five of the neutrino mixing parameters, three mixing angles $\theta_{12}$, $\theta_{13}$, $\theta_{23}$, and the two squared-mass differences $\Delta m_{21}^2$, $|\Delta m_{32}^2|$ up to 3$\sigma$ confidence level~\cite{DayaBay:2022orm, NOvA:2017ohq, T2K:2018rhz, Super-Kamiokande:2017yvm, NOvA:2019cyt}. Among these parameters, the sign of atmospheric mass-squared difference,\rm{i.e.} $\Delta m_{31}$, which will determine the mass ordering problem, the octant of the mixing angle  $\theta_{23}$, and the the true value of the CP violating phase $\delta_{CP}$ are left to be unknown. The determination of the CP-violating phase, the Dirac phase, has been the core research program in neutrino physics for years because it provides a potential source of CP violation in the SM lepton sector. It has been known that the leptonic CP violation could generate the matter-antimatter asymmetry through leptogenesis~\cite{Fukugita:1986hr}.
CP violation in neutrino oscillation can be measured through the difference between the oscillation probability of the neutrino and antineutrino, expressed as $\deltaP = P_{\alpha\beta} - \overline{P}_{\alpha\beta}$, which is well quantified by $\delta_{\rm CP}$. There are several experiments worldwide dedicated to the measurements of the neutrino parameters, especially the CP phase, performing searches of short-baseline and long-baseline neutrino oscillation. To ensure that there are enough neutrino flavors oscillated from source neutrino and to be detected by Far Detector (FD), a long-baseline neutrino oscillation experiment is preferable rather than a short-baseline. Recently, the long-baseline experiments, T2K (Tokai to Kamioka)~\cite{T2K:2018rhz, T2K:2019bcf, T2K:2021xwb, T2K:2023smv,Walsh:2022pqg} and  NOvA~\cite{NOvA:2021nfi} report their results.  T2K reports a measured value for CP phase, $\delta_{\rm CP} = -1.97_{-0.70}^{+0.97}$ while excluding $\delta_{\rm CP}$ = 0 and $\pi$ at 90\% CL, indicating CP violation in the lepton sector at relatively improved confidence level. However, there is CP conserving values for $\dcp$ within $3\sigma$ standard error~\cite{Walsh:2022pqg}. The FD in this case is the Super-Kamiokande, a 50 Kton water Cherenkov detector. A narrow band neutrino beam is produced at an angle of $2.5^{\circ}$ by a 30 GeV proton beam hitting on graphite target. With this off-axis method, the narrow band neutrino energy has a peak at 0.6 GeV. The secondary neutrino produced from decays of Kaon or Pion travels a distance of 295 Km to reach the Super-Kamiokande detector.
T2K plans to extend its term to 2026, followed by the Hyper-K project~\cite{Hyper-Kamiokande:2018ofw} with the mass of the far detector to be increased by a factor of 10, and will offer a broad science program. On the other hand, the NOvA experiment~\cite{NOvA:2021nfi} is also a long-baseline accelerator-based neutrino oscillation experiment. It uses the upgraded Fermilab NuMI beam and measures electron-neutrino appearance and muon-neutrino disappearance at its far detector in Ash River, Minnesota. The reported NOvA result shows no strong preference for any particular value of the neutrino CP phase within Normal Ordering (NO) and has a visible tension with T2K's measurement while agrees at 90\% confidence level in Inverted Ordering (IO). This slight tension may arise because of the systematic uncertainties or it is probably a hint for new physics effects arising from sterile neutrinos or non-standard neutrino interactions~\cite{Abazajian:2012ys, Palazzo:2013me, Giunti:2019aiy,Kopp:2013vaa,Ohlsson:2012kf,Farzan:2017xzy}.

Another promising long-baseline neutrino experiment under construction is DUNE (Deep Underground Neutrino Experiment)~\cite{DUNE:2015lol, DUNE:2020mra, DUNE:2020ypp,DUNE:2020txw, DUNE:2021cuw,DUNE:2020jqi}, whose goals are the determination of the neutrino mass ordering, observation of CP violation (up to 50\% level), and precise measurements of oscillation parameters, such as $\delta_{\rm CP}$, $\sin^2(2\theta_{21})$. The idea is to send a wide-band high-intensity muon neutrino beam from Fermilab to the Sanford Underground Facility in Homestake at the 1300 Km distance. The detector technology of DUNE experiment is based on building liquid argon time projection chambers (LArTPC). Unlike the T2K experiment, the neutrino beam energy has a peak at 2.5 GeV with a broad range of neutrino energies. The neutrino beam is produced from proton collision on the graphite target. In the corresponding DUNE TDR report~\cite{DUNE:2020lwj,DUNE:2020ypp,DUNE:2020mra,DUNE:2020txw}, it is shown that favorable values for $\delta_{\rm CP}$ with $3\sigma$ $(5\sigma)$ can be achieved after five (ten) years of running. It is worth noting that although muon beams produced from proton-on-target experiments like T2K, NOvA, and DUNE can have higher luminosity and energy, but there are also some disadvantages \rm{i.e.} beam contamination because of the intermediate hadronic states and their decay products, higher emittance etc. It is also worth to pay attention that produced neutrino and antineutrino beams in the above experiments can not be run at the same time, which takes much longer time than a simultaneous run of neutrino and anti-neutrino beams.

This motivates us to examine the physics potential of muon beams produced from lepton collider, especially positron-on target experiment. In this letter, we are interested in applying collimated muon beams into neutrino mixing and CP phase measurements. Although the beam density is lower than the proton-on-target scenario, there are several significant benefits leading to large neutrino flux and high sensitivity on CP phase, including 1) collimated and manipulable muon beams, which lead to a larger acceptance of neutrino sources in the far detector side; 2) symmetric $\mu^+$ and $\mu^-$ beams, and thus symmetric neutrino and antineutrino sources, which make this proposal ideally suitable for measuring neutrino CP violating phase. Importantly, antineutrino and neutrino flux distributions produced collimated muon beams are same, and thus for example, $\bar{\nu}_{e}\rightarrow\bar{\nu}_\mu$ and $\nu_{e}\rightarrow \nu_\mu$ oscillation signals can be collected simultaneously, without  needs for separate runs for neutrinos or antineutrinos, as usually done in the other long baseline neutrino experiments.

As to be discussed below, the estimated neutrino flux in our proposal is comparable to or even larger than the DUNE experiment. The neutrino energy has wide distributions in 1-20 GeV region, and peaks at around 5-15 GeV (neutrino energy can be further tuned with on-axis and off-axis techniques), suggesting our proposal is also suitable for tau neutrino studies, because the peaked energy is much greater than the threshold energy for tau lepton production. Taking into account both muon and electron neutrinos and antineutrinos, the signal yields indeed can be doubled or more. Finally, we point out that it is possible to exchange $\mu^+$ and $\mu^-$ flying routes, and consequently, reducing possible bias or systematic uncertainties.  

We perform a prospective study mainly dealing with the CP violation sensitivities using \globes, an open source program for simulating standard or non-standard neutrino oscillations including matter effects~\cite{Wolfenstein:1977ue,  Huber:2004ka, Huber:2007ji}. This software is kept tested by super-beam neutrino oscillation experiments over several decades. The outline of the paper is organized as follows. In Sec.~\ref{sec:theory}, we describe some relevant theory for neutrino and anti-neutrino oscillation probability and show numerical results as well as the description of the experimental setup for our proposal. The results of simulation based on our proposal are discussed in Sec.~\ref{sec:results}. There is a little discussion in Sec.~\ref{sec:nues}about the physics potential of our proposal for studying neutrino oscillations when including sterile neutrinos. Finally, Sec.~\ref{sec:summary} gives a conclusion and future outlook in the end.

\section{ Theory and Experimental setup} \label{sec:theory}
\subsection{Oscillation probability}
\begin{figure}[h]
    \centering
    \includegraphics[width=.45\columnwidth]{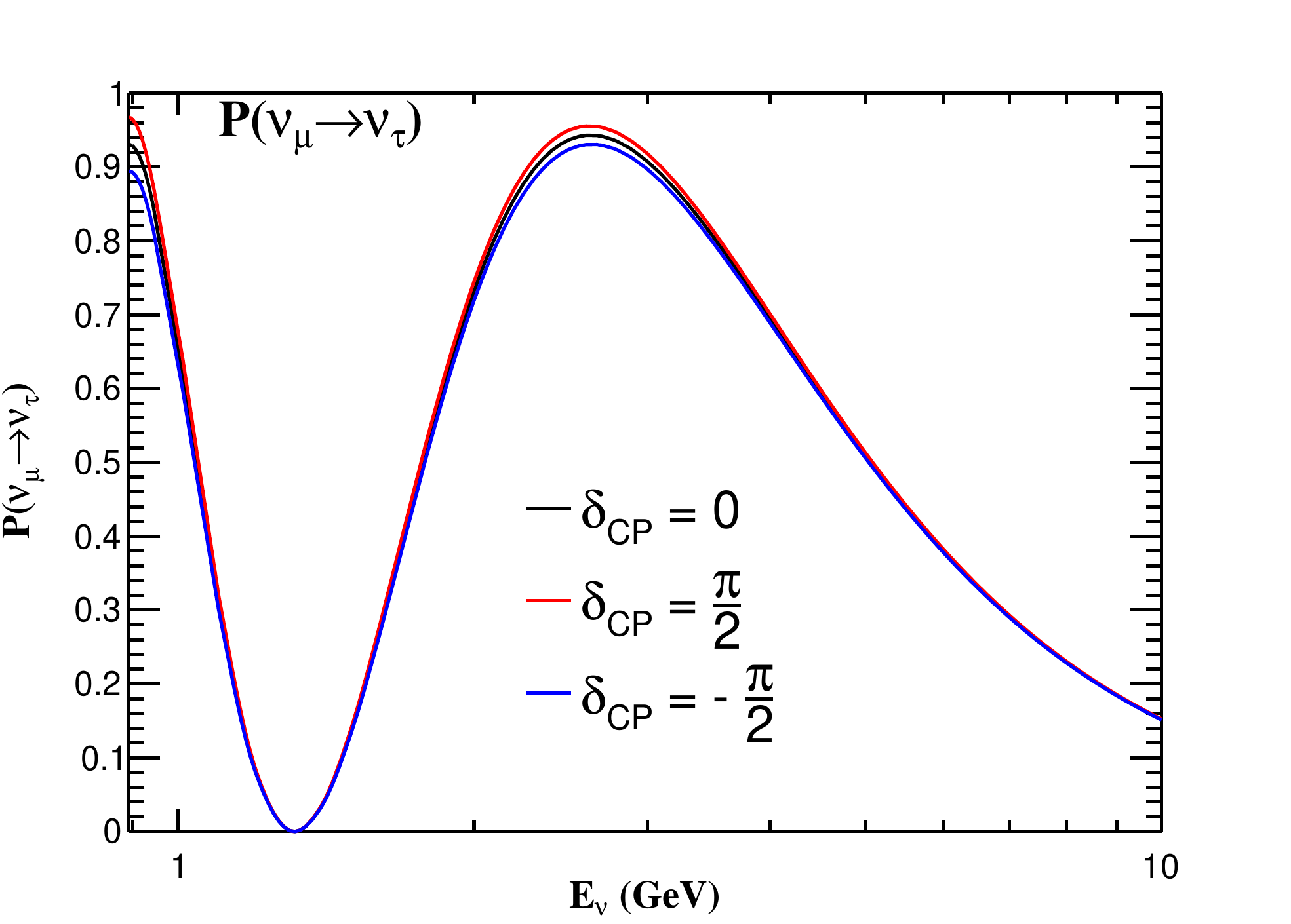}
    \includegraphics[width=.45\columnwidth]{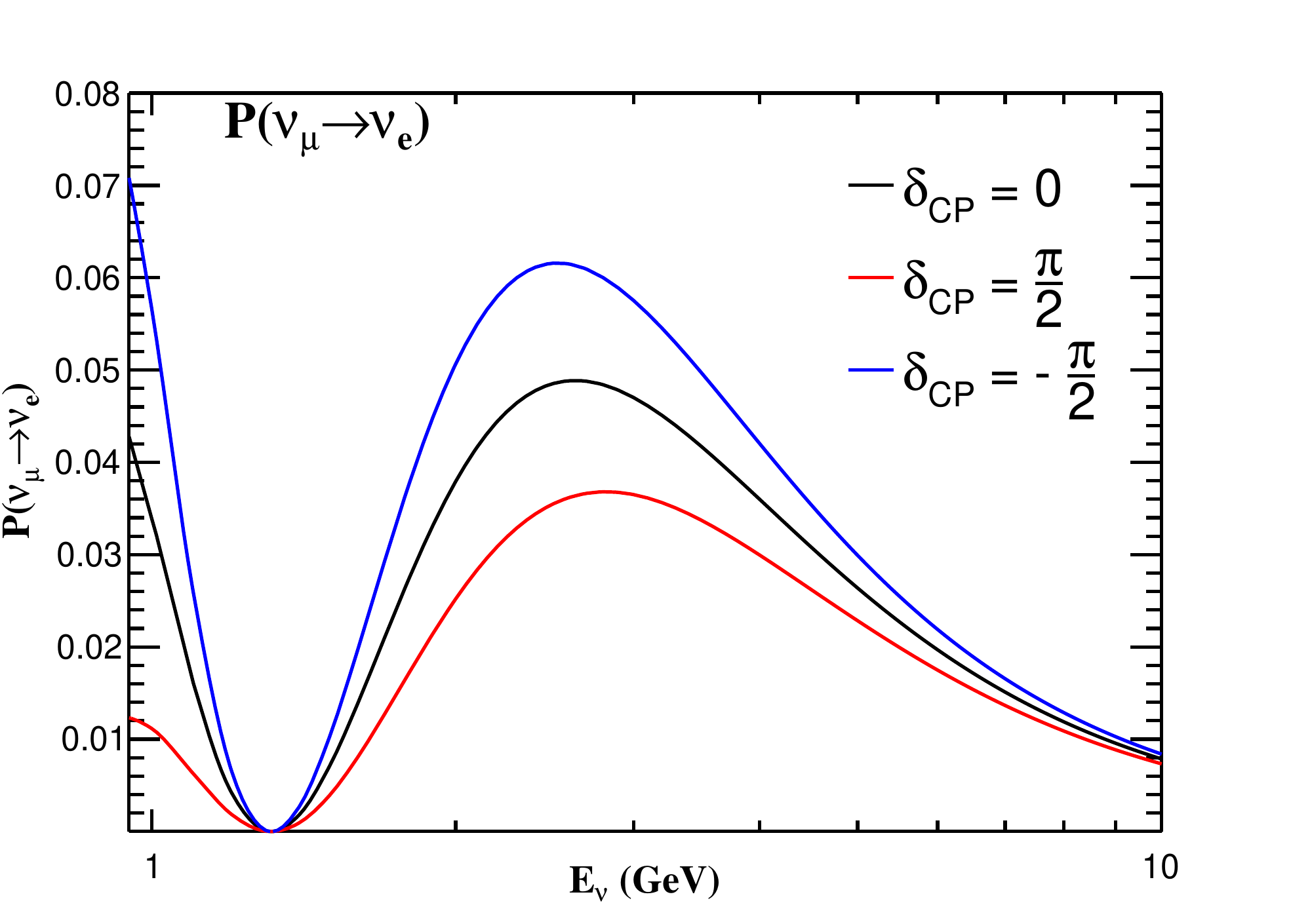}
    \includegraphics[width=.45\columnwidth]{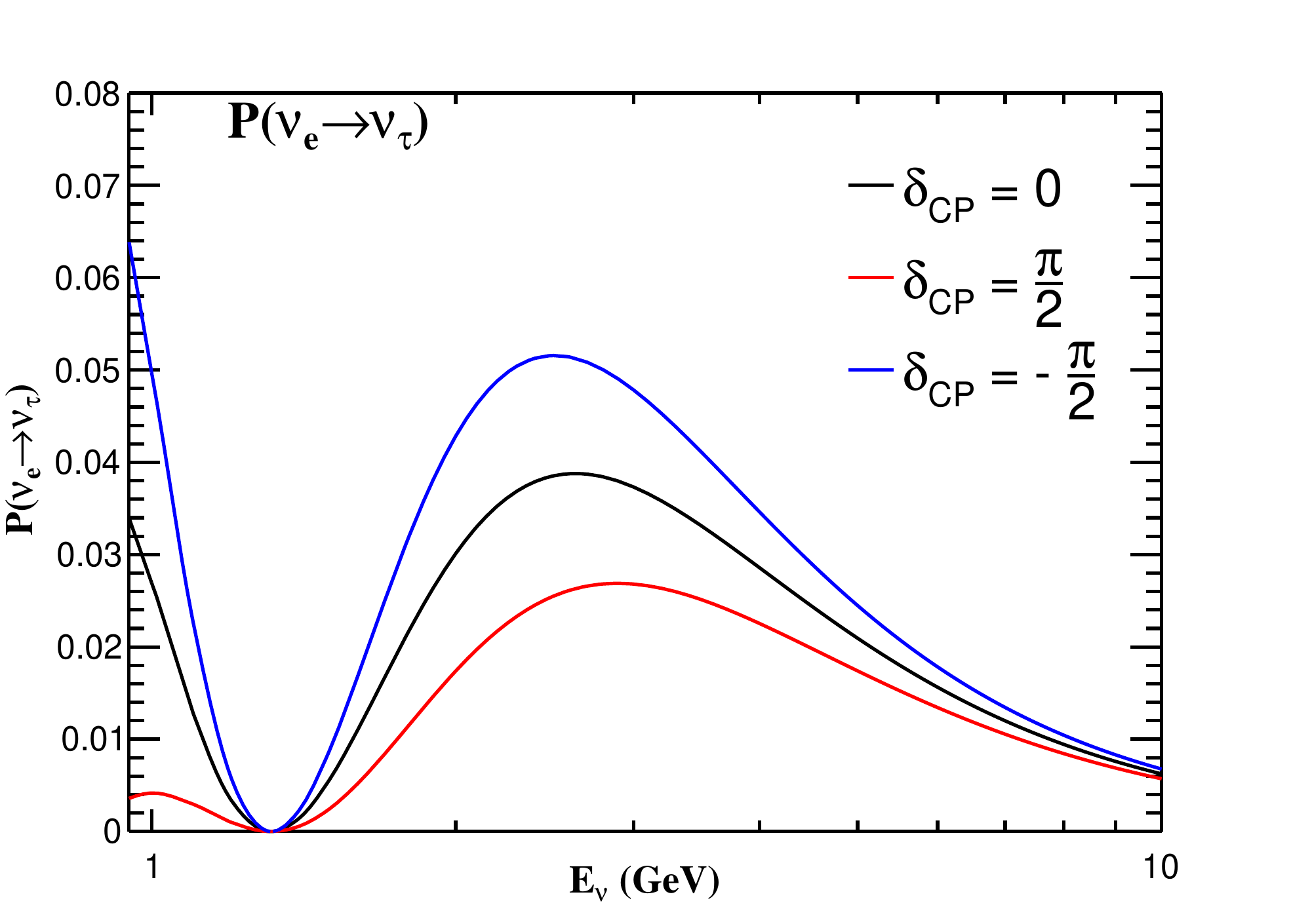}
    \includegraphics[width=.45\columnwidth]{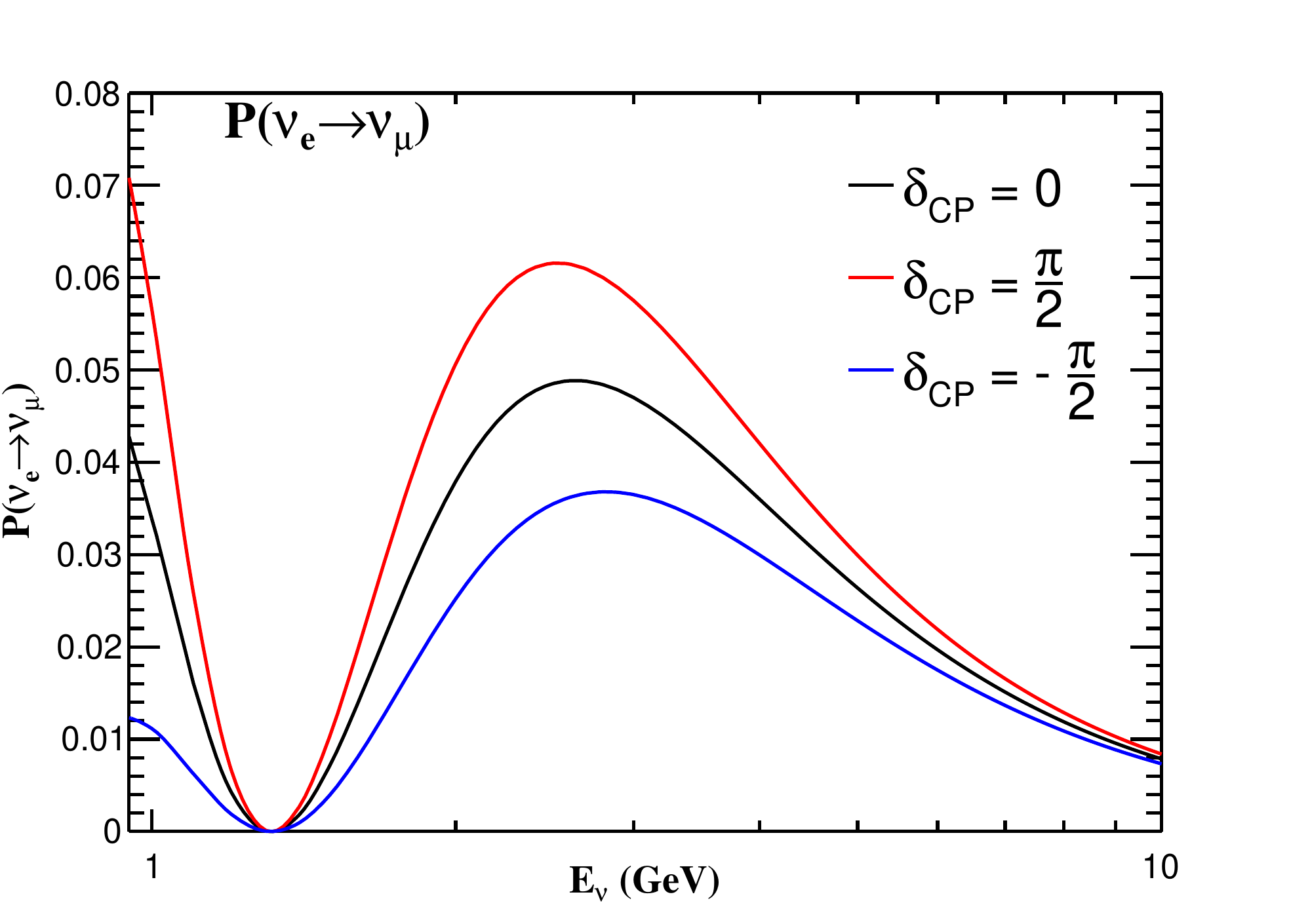}
    \caption{The vacuum oscillation probability for $\nu_{\mu}\to\nu_{\tau}$, $\nu_{\mu}\to\nu_e$, $\nu_e\to\nu_{\tau}$, and $\nu_e\to\nu_\tau$ transitions at three different $\delta_{\rm CP}$ values as function of $\nu_{\mu}$ energy. The length of oscillation baseline is set to L =1300 Km.}
    \label{fig:vacosc}
\end{figure}

Here we discuss some interesting properties of neutrino oscillation probability with regard to the energy and baseline length. We start with a general discussion regarding standard oscillation probabilities.
Based on the general knowledge of quantum mechanics, we can obtain the transition amplitude as inner product of initial and final neutrino flavor states~\cite{Nunokawa:2007qh}. We derive appearance probability for $\nu_{\tau}$, $\nu_{\mu}$, $\nu_e$ and their anti neutrinos following the general formula~\ref{eq:osc}
\begin{equation}
    P(\nu_{\alpha}\rightarrow\nu_{\beta}) = \delta_{\alpha\beta} -4 \sum_{i< j}^n {\rm Re} \lb U_{\alpha i}U^*_{\beta i}U^*_{\alpha j}U_{\beta j}\sin^2X_{ij}\rb + 2 \sum_{i< j}^n {\rm Im} \lb U_{\alpha i}U^*_{\beta i}U^*_{\alpha j}U_{\beta j}\rb \sin2X_{ij},
    \label{eq:osc}
\end{equation}
where $U_{\alpha i}$ are the elements of PMNS mixing matrix, and $X_{ij}$ reads as
\begin{equation}
    X_{ij} = \frac{(m_i^2 - m_j^2)L}{4E_{\nu}} = 1.267 \frac{\Delta m_{ij}^2}{\rm eV^2} \frac{L}{\rm Km} \frac{\GeV}{E_{\nu}},
\end{equation}
$L$ is the length of the neutrino propagation distance whose unit is ${\rm Km}$ and $E_{\nu}$ corresponds to the true neutrino energy in ${\rm GeV}$ unit, $\Delta m_{ij}$ is the squared mass difference in ${\rm eV^2}$ unit.
The oscillation probabilities for antineutrino can be obtained with the exchange $U\rightarrow U^*$. The first term in Eq.~\ref{eq:osc} is CP conserving since it is the same for neutrinos and anitneutrinos, while the last one is CP violating because it has opposite signs for neutrinos and antineutrinos.  As mentioned in the introduction, the simultaneous run of $\mu^+$ and $\mu^-$ beams will provide eight oscillation channels through the decay $\mu^-\rightarrow \nu_{\mu}+\overline{\nu}_e$ and $\mu^+\rightarrow\overline{\nu}_{\mu} + \nu_e$:
\begin{align*}
    &\nu_{\mu} \rightarrow \nu_{\tau},~~~\overline{\nu}_{\mu} \rightarrow \overline{\nu}_{\tau} \\
    &\nu_{\mu} \rightarrow \nu_{e}, ~~~\overline{\nu}_{\mu} \rightarrow \overline{\nu}_{e}\\
    &\nu_{e} \rightarrow \nu_{\tau}, ~~~\overline{\nu}_{e} \rightarrow \overline{\nu}_{\tau}\\
    &\nu_{e} \rightarrow \nu_{\mu}, ~~~\overline{\nu}_{e} \rightarrow \overline{\nu}_{\mu}
\end{align*}
Based on equation~\ref{eq:osc}, we can get some approximated the vacuum oscillation probabilities for the above eight oscillation channels as
\begin{subequations}
\begin{align}
P(\nu_\mu \to \nu_\tau) &\simeq \sin^2\left(2\theta_{23}\right)\cos^4(\theta_{13})\sin^2\left(1.27\frac{\Delta m_{32}^2 L}{E_{\nu}}\right) \pm 1.27\Delta m_{21}^2\frac{L}{E_{\nu}}\sin^2\left(1.27\frac{\Delta m_{32}^2L}{E_{\nu}}\right) \times 8 J_{\rm CP}, \label{eq:pmutau} \\ 
P(\nu_{\mu}  \rightarrow \nu_{e} ) &\simeq  \sin^2(2\theta_{13})\sin^2(\theta_{23})\sin^2\left(1.27\Delta m^2_{32}\frac{L}{E_{\nu}}\right)\mp 1.27\Delta m^2_{21}\frac{L}{E_{\nu}}  \sin^2 \left(1.27\Delta m^2_{32}\frac{L}{E_{\nu}}\right)\times 8J_{\textrm{CP}},\label{eq:pmuel} \\ 
P(\nu_{e}  \rightarrow \nu_{\tau} ) &\simeq  \sin^2(2\theta_{13})\cos^2(\theta_{23})\sin^2\left(1.27\Delta m^2_{32}\frac{L}{E_{\nu}}\right)\mp 1.27\Delta m^2_{21}\frac{L}{E_{\nu}} \sin^2 \left(1.27\Delta m^2_{32}\frac{L}{E_{\nu}}\right) \times 8J_{\textrm{CP}},\label{eq:peltau}  \\ 
P(\nu_{e}  \rightarrow \nu_{\mu} ) &\simeq  \sin^2(2\theta_{13})\sin^2(\theta_{23})\sin^2\left(1.27\Delta m^2_{32}\frac{L}{E_{\nu}}\right)\pm 1.27\Delta m^2_{21}\frac{L}{E_{\nu}}  \sin^2 \left(1.27\Delta m^2_{32}\frac{L}{E_{\nu}}\right) \times 8J_{\textrm{CP}}, \label{eq:pelmu}
\end{align}
\label{eq:oscp}
\end{subequations}
where $J_{\rm CP}$ is the ``Jarlskog invariant''~\cite{Jarlskog:1985ht} and replacing the mixing parameters with up-to-date measured values~\cite{PDG:2020ssz},
\begin{equation}\begin{split}
\label{eqn:JCP}
J_{\mathrm{CP}}&\equiv\sin\theta_{13}\cos^2\theta_{13}
\sin\theta_{12}\cos\theta_{12}
\sin\theta_{23}\cos\theta_{23}
\sin\delta_{\textrm{CP}}\\
 &= 0.03359 \pm 0.0006 (\pm0.0019)\sin\delta_{\textrm{CP}}.
\end{split}
\end{equation}
It should be pointed out that we drop the sub-leading terms, second order in $\sin\theta_{13}$ and $\Delta m_{21}^2$ in the above oscillation probabilities. The above form of the oscillation probabilities are applicable for most of the super beam experiments without the consideration of the matter effects. However, this form of oscillations might differ from the results obtained with \globes. The vacuum oscillation probabilities in Eq.~\ref{eq:oscp} are plotted in Fig.~\ref{fig:vacosc} 
with respect to the neutrino energy in the $1 \GeV < E_{\nu} < 10 \GeV$ range. Here the probability for anti-neutrino oscillations can be obtained by simply replacing $\delta_{\rm CP}$ with $-\delta_{\rm CP}$.
The oscillation probability difference between $P(\nu_{\alpha}\rightarrow\nu_{\beta})$ and $P(\bar{\nu}_{\alpha}\rightarrow\bar{\nu}_{\beta})$ reads as
\begin{equation}
    \Delta P(\nu_{\alpha}\rightarrow\nu_{\beta})= 16J_{\rm CP}\times 1.27\Delta m_{21}^2\frac{L}{E_{\nu}}\sin^2\left( \frac{\Delta m_{32}^2L}{E_{\nu}} \right)
\end{equation}

Using the current measured values of the mixing angles and squared mass differences~\cite{PDG:2020ssz} and taking the distance of neutrino propagation as L =1300 Km, we have the numeric values for the neutrino oscillations at $E_{\nu} = 7\, (5)~\rm{GeV}$ as
\begin{subequations}
\begin{align}
\pmutau &=0.2916 \pm 0.0026\sin\delta_{\rm CP} \,(0.5093 \pm 0.0048\sin\delta_{\rm CP}),\\
\pmuel & =0.0151 \mp 0.0026\sin\delta_{\rm CP} \,(0.0264 \mp 0.0048\sin\delta_{\rm CP}),\\
\pelmu & =0.0151 \pm 0.0026\sin\delta_{\rm CP} \,(0.0264 \pm 0.0048\sin\delta_{\rm CP}),\\
\peltau& =0.0119 \mp 0.0026\sin\delta_{\rm CP} \,(0.0209 \mp 0.0048\sin\delta_{\rm CP}).
\end{align}
\label{eq:ocp}
\end{subequations}

The impact of matter effects on the appearance probability may be negligible in the short baseline oscillation experiments and for reactor neutrinos. Given the significant length of the oscillation baseline in our proposal, matter effects arising from the interaction of neutrinos with nucleons of matter as they propagate through the Earth's internal crust should be taken seriously, because the matter-induced correction factor for the oscillation probability is relatively larger than other short baseline oscillation experiments~\cite{Agarwalla:2013tza,Sharma:2023jzg}. This may bring some visible changes to our vacuum oscillation probabilities shown in the Fig.~\ref{fig:vacosc}. A complete set of neutrino oscillation probabilities including matter effect with constant density for three flavors are obtained through series expansions in the mass hierarchy parameter, $\Delta m_{21}^2/\Delta m_{31}^2$, and mixing parameter $\sin\theta_{13}$ with first and second order, which is available  in~\cite{Akhmedov:2004ny}. 
\begin{figure}[ht]
    \centering
    \includegraphics[scale=0.4]{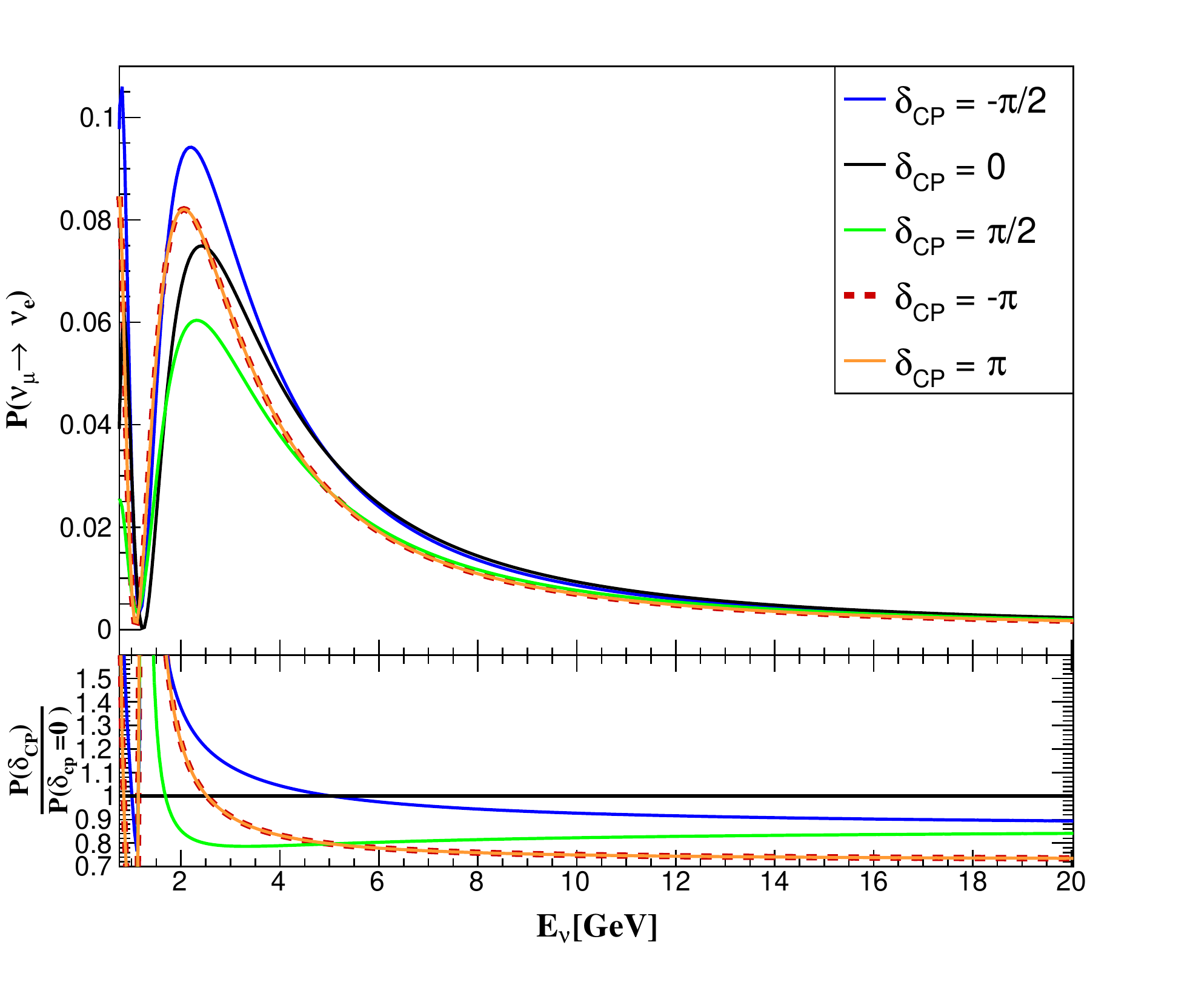}
    \includegraphics[scale=0.4]{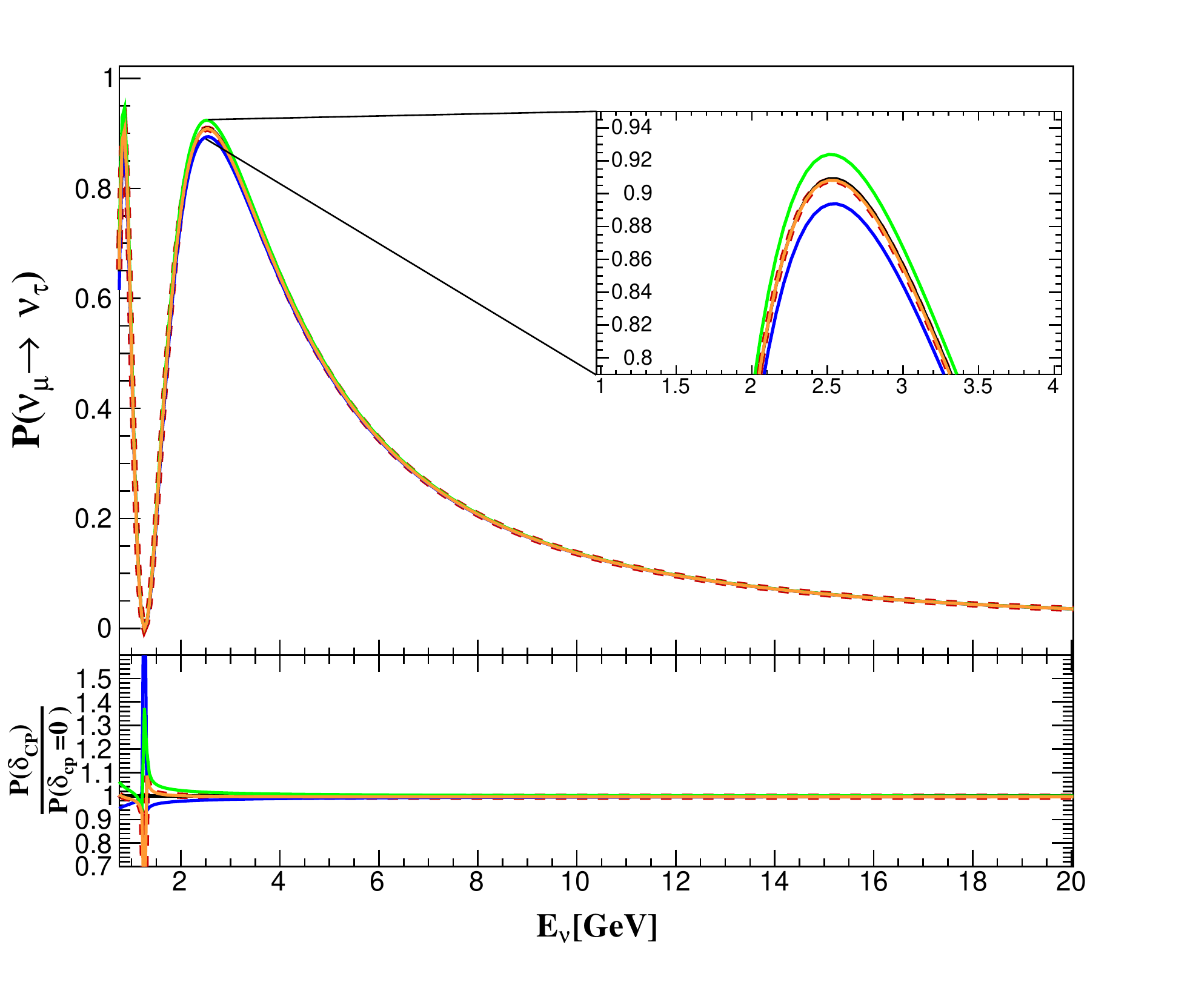}
    \includegraphics[scale=0.4]{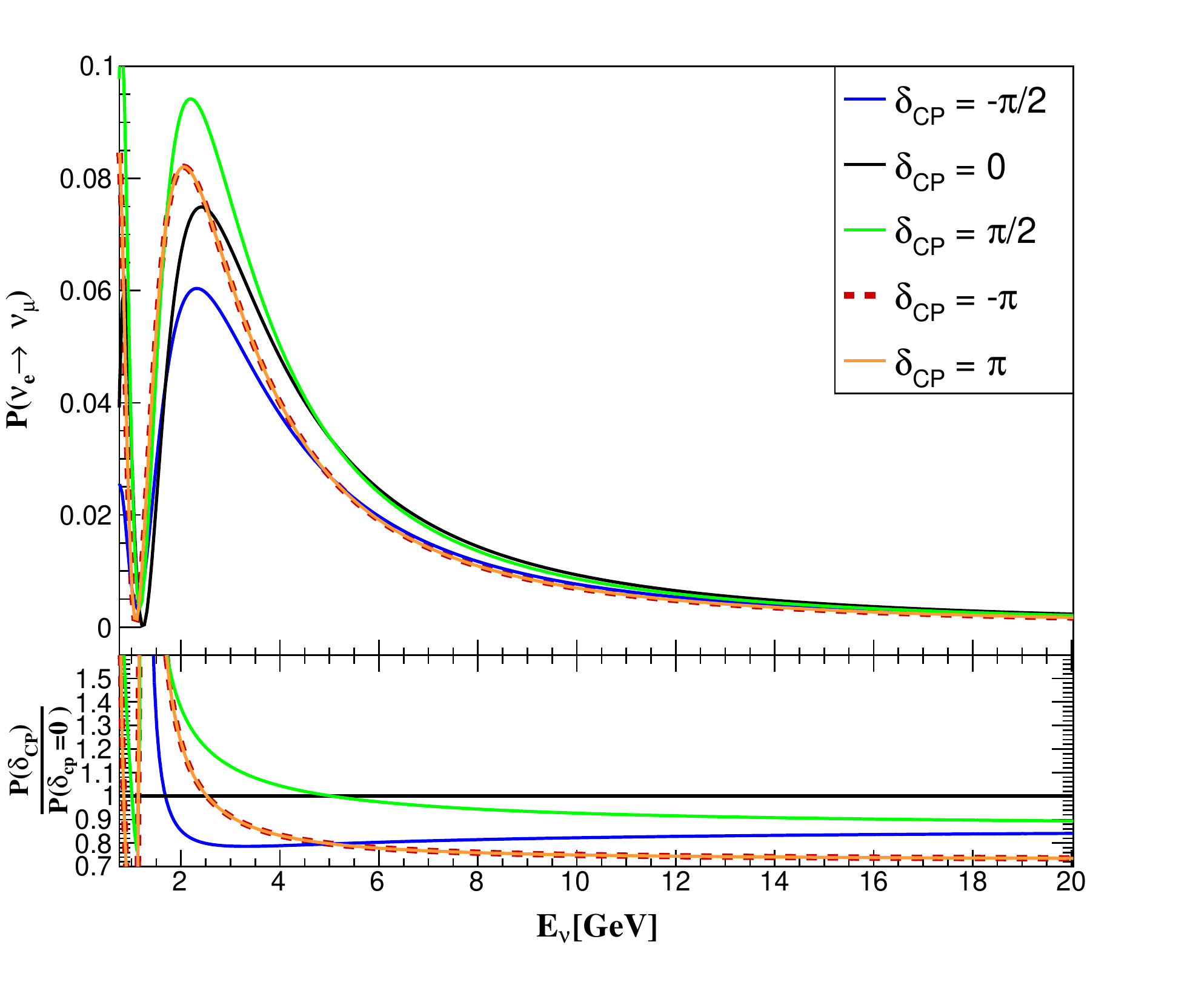}
    \includegraphics[scale=0.4]{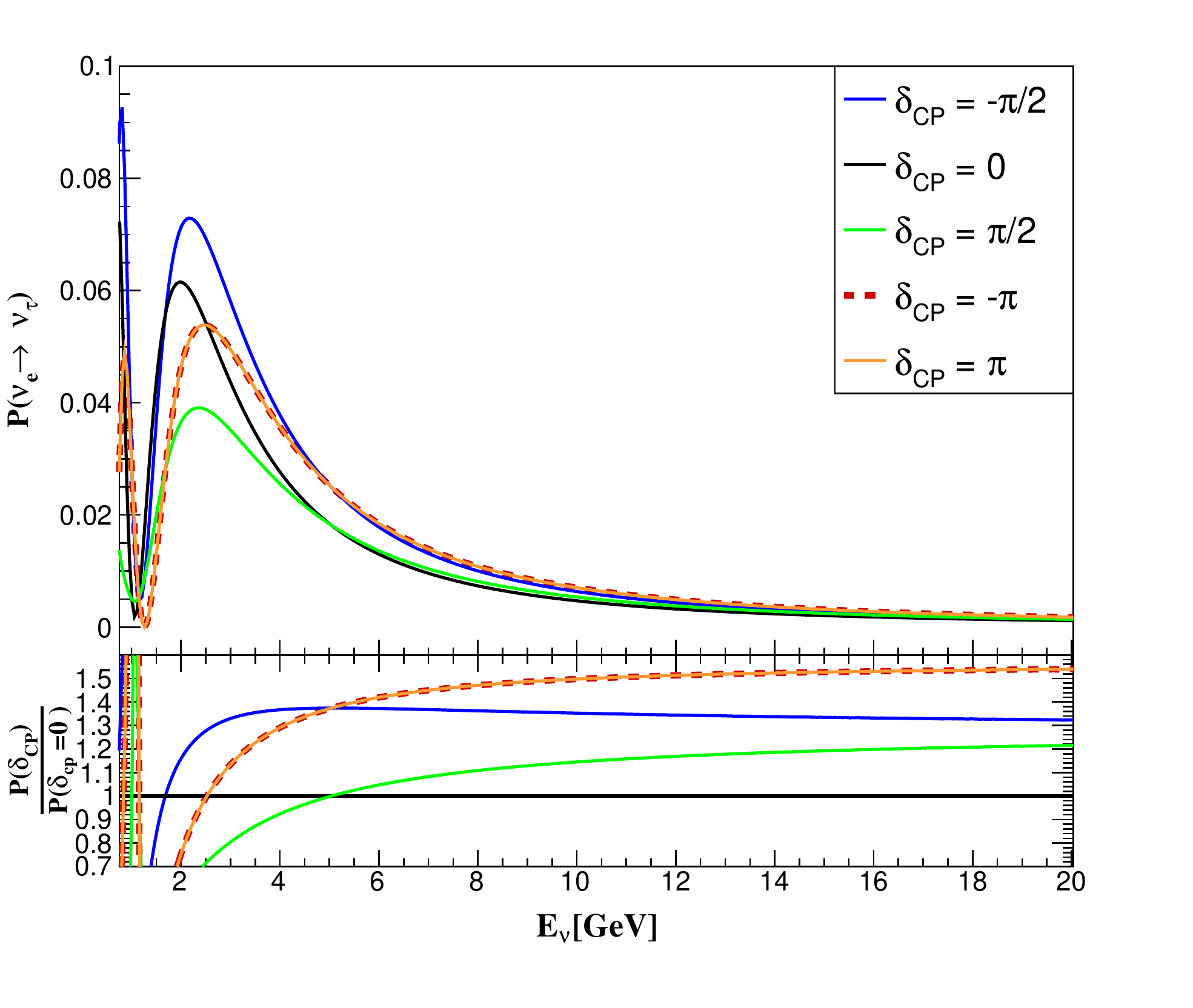}
    \caption{Oscillation probability for $\pmuel$ (top left), $\pmutau$ (top right), $\pelmu$ (bottom left), and $\peltau$ (bottom right) oscillation channel. Transitions are the function of the neutrino reconstructed energy and depicted for five $\dcp$ values: $\dcp = -\pi/2$, 0, $\pi/2$, $-\pi$, $\pi$. In each plot, the ratio $P(\dcp\neq0)/P(\dcp = 0)$ is shown in the bottom panel.
    The oscillation baseline length is taken as $L = 1300$ Km, and Normal Mass ordering is assumed. For the other oscillation parameters, we use the values given in Tab.~\ref{tab:params}.}
    \label{fig:opMatter}
\end{figure} 

Now we demonstrate the appearance probabilities for $\pmuel$, $\pmutau$, $\pelmu$, $\peltau$ channel that depict the impact of matter effects on the neutrino oscillation. The probability is the function of the reconstructed neutrino energy and drawn for five fixed $\dcp$ values as shown in the Fig.~~\ref{fig:opMatter}. The overall size of the probabilities are almost the same for $\pmuel, \pelmu, \peltau$ in each plot except for $\pmutau$ oscillation. The transition probability for $\pmutau$, shown on the top-right plot, is much greater than the other three probabilities thus enhances the $\nu_{\tau}$ neutrino-related events at the far detector.
From the ratio of probabilities with non-zero $\dcp$ to probabilities with $\dcp=0$, $P(\dcp\neq0)/P(\dcp =0)$, shown at the bottom of each figure, it is very clear that the inclusion of matter effects change the shapes of oscillation probability significantly relative to the vacuum oscillations displayed in Fig.~\ref{fig:vacosc}. Two of the five $\dcp$ values, $-\pi/2$ (blue line) and $\pi/2$ (green line), maximize the CP violation the most while the other three values are thought to be CP conserving phase: $\pi$ (orange line), $-\pi$ (red dashed line), $0$ (black line). Therefor oscillation probability is supposed to be the same for these three CP conserving $\dcp $ values. However, there is a difference between them: $P(\dcp = 0)\neq P(\dcp =-\pi) = P(\dcp = \pi)$. This may arise from  the collective influence of the matter effects and the expansion method. It is worth noting that, for the oscillation channels $\pmuel$ and $\pelmu$, the probability with $\dcp=0$ is larger than that of the other $\dcp$ values. As for the tau neutrino appearance from $\nu_{\mu}$ and $\nu_e$ channels, the outcome is different. In the  $\nu_{\mu}\rightarrow\nu_{\tau}$ oscillation channel, the difference of the oscillation probability for CP-conserving $\dcp$ values is very small or even become zero as the neutrino energy increases. We will see further outcome with both the event spectrum and significance in the following sections.

\begin{figure}[h]
    \centering
    \includegraphics[width=.9\columnwidth]{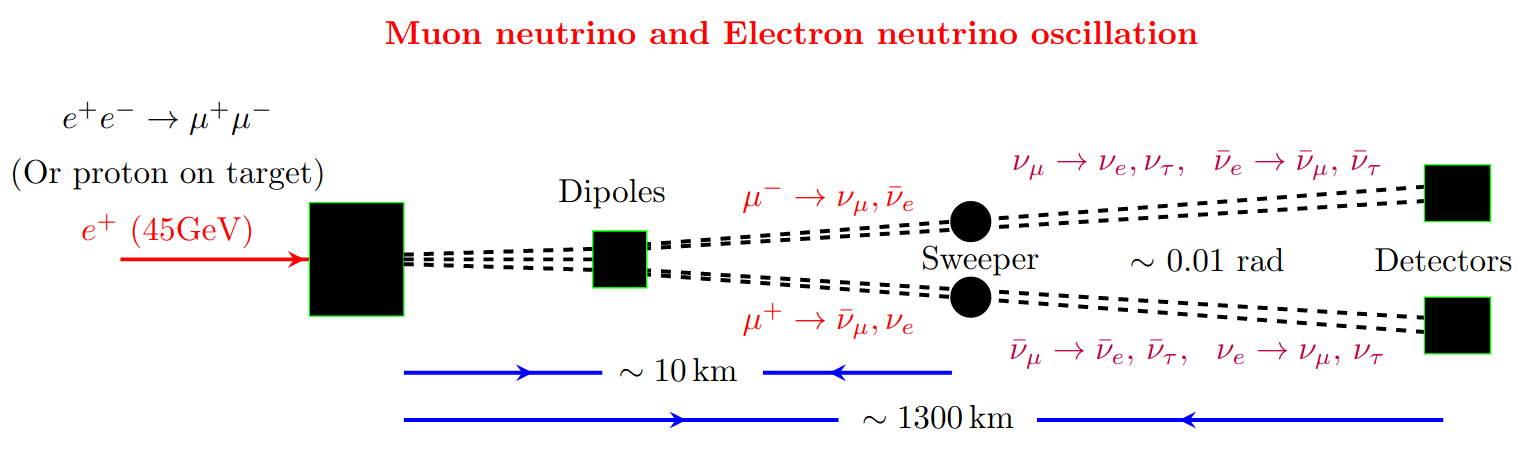}
    \caption{A proposed neutrino oscillation experiment to probe neutrino violation CP phase by measuring muon electron (anti-) neutrino oscillation and the differences of resulted $\nu_{e,\mu}\rightarrow\nu_\tau$, $\nu_e\rightarrow\nu_\mu$, and their antineutrino correspondents. This proposal is based on collimated muon beams achieved from e.g., Positron on Target method, where 45 GeV positron beams are fired. Dipoles are used to separate $\mu^+$ and $\mu^-$ with an angle around 0.01 rad., with direction changeable. Muon beams fly about 10 Km and radiate neutrinos before being swept away. Neutrinos then further fly 1300 Km to reach DUNE type of detectors.}
    \label{fig:exp}
\end{figure}
\subsection{Experimental setup and qualitative estimations}

Fig.~\ref{fig:exp} shows a proposed neutrino oscillation experiment to probe neutrino CP phase by measuring muon electron (anti-)neutrino mixing and their differences. We are especially interested in the oscillation modes of $\nu_{\mu}\rightarrow\nu_{e, \tau}$, $\nu_e\rightarrow\nu_{\mu, \tau}$ and their corresponding antineutrinos with more details to be given later. This proposal is based on collimated muon beams achieved from e.g., Positron on Target method, where 45 GeV positron beams are shed on the target. Dipoles are used to separate $\mu^+$ and $\mu^-$ with an angle around 0.01 rad., with direction changeable (notice the acceptance of the dipole to separate muon and anti-muons is not considered here and needs to be evaluated later). Muon beams fly about 10 Km and radiate neutrinos before being swept away. Neutrinos then further fly e.g., 1300 Km to reach DUNE or T2K type of detectors~\cite{DUNE:2020mra}. 

\begin{figure}[ht]
    \centering
    \includegraphics[width=.45\columnwidth]{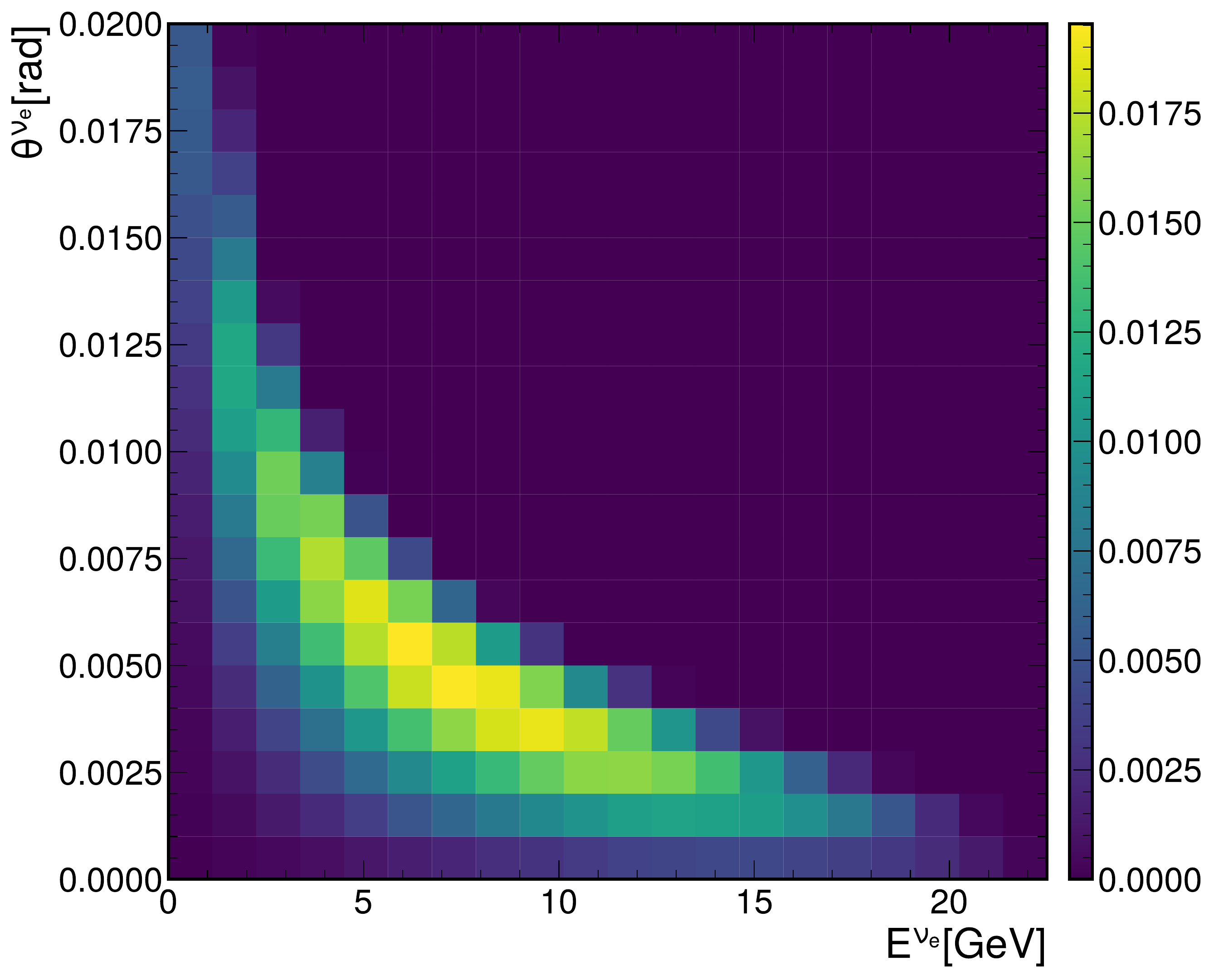}
    \includegraphics[width=.45\columnwidth]{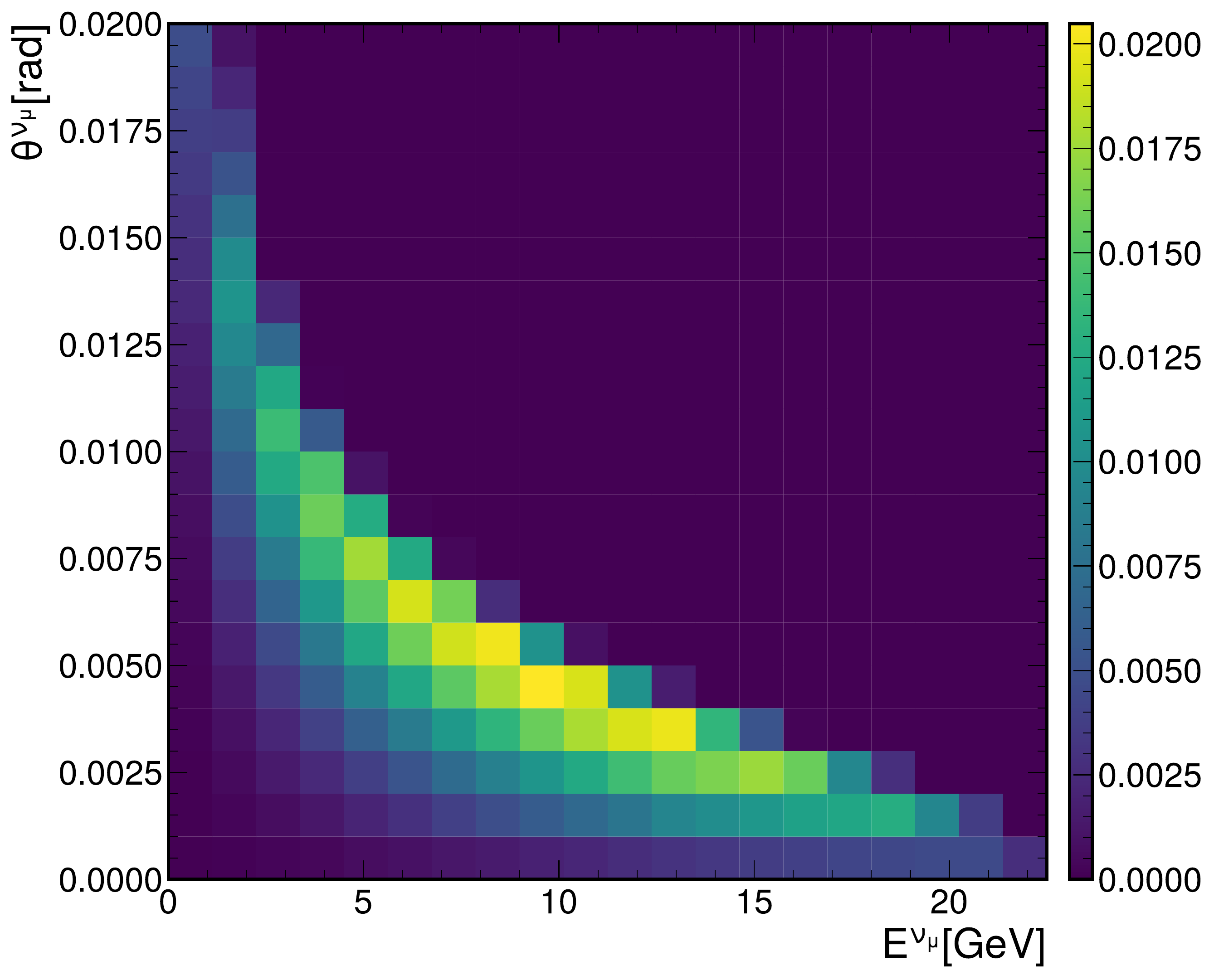}
    \caption{2D distributions of energy and angle in respect to muon flying direction, for muon and electron neutrinos from 22.5 GeV $\mu^+\rightarrow e^+\,\bar{\nu}_{\mu}\, \nu_{e}$ (similarly for $\mu^-$ decay).}
    \label{fig:neu2D}
\end{figure}

 \begin{itemize}
     \item Muon production rates. As estimated in the formula~\ref{eq:murate}, the produced muon numbers is
     $n(\mu^+ \mu^-)_{max}\approx n^+\times 10^{-5}$~\cite{Alesini:2019tlf}. Assuming positron bunch density as $10^{12}$/bunch and bunch crossing frequency as $10^5$/sec, we get {\bf muon production rates $dN_\mu/dt \sim 10^{12}$/sec (or $10^{19}$/year)}. 
     
     Notice although we take positiron on target muon source as an example, proton on target muon beams should also be applicable. Moreover, the future TeV scale muon collider is indeed targeting a much larger intensity beam by more than 1-2 orders of magnitudes~\cite{MuonCollider:2022nsa,MuonCollider:2022glg, MuonCollider:2022xlm}.
     \item For muons with the energy around 20 GeV, the mean flying distance is around 100 Km. If there is a straight tube with a length around 5-10 Km to let muons go through with quadrupoles to keep them merged (To further reduce the angular emittance with quadrupoles is to be checked), {\bf the decayed fraction can reach $10^{-1}$} in realistic case. On the other hand, we can also refer to a muon complex as discussed in Ref.~\cite{Quigg:1997uk}, where the muon beam is accelerated in a circular section and then extracted into the rectangular section for decays. The intensity of the neutrino beam compared with the incoming muon beam is suppressed by a ratio around $10^{-1}$, i.e., the fraction of the collider ring circumference occupied by the production straight section.
     \item The opening angles between muon beams axis and the momentum of decay products are around 0.005 rad. as shown in Fig.~\ref{fig:neu2D} and may be kept smaller with quadrupoles. For neutrinos traveling 1300 Km to reach far detectors, the spread size can be around 1-5 Km. For a DUNE-like detector with a cubic size of about 20 m~\cite{DUNE:2020mra},  {\bf the neutrino acceptance is then $10^{-4}$}.
     \item Neutrino and anti-neutrino interactions inside detectors. With a $L=20$ m long liquid Argon detector (DUNE far detector indeed has a length around 50m~\cite{DUNE:2020mra}), {\bf the expected event yield rate can be roughly estimated with: $dN_\mu/dt\times L \times\sigma_{n\nu}\times \rho N_A\cdot dE$}, where $N_A$ is the Avogadro constant, $\rho\sim $ 2.834 $\rm{g}/\rm{cm}^3$, $\sigma_{n\nu}$ represents the neutrino-nucleon cross sections and is approximately {\bf $10^{-38} \rm{cm}^2$} for a 10 GeV neutrino~\cite{PDG:2020ssz,Formaggio:2012cpf}. Actually, the cross sections are function of the neutrino energy energy. We will show the simulated results for neutrino cross sections in below sections. 
 \end{itemize}

Because of its heavy mass and very short lifetime, the tau neutrino production with abundant numbers in conventional accelerator is very difficult. On the contrary, we have rich tau neutrino flux because of the higher $\pmutau$ oscillation shown in top right corner of Fig.~\ref{fig:vacosc}. The tau neutrino flux can be further strengthened if we consider the oscillation channel of $\peltau$ whose oscillation probability is much more smaller than $\pmutau$. Considering the fraction of muon neutrino oscillated into tau neutrino~\cite{DeGouvea:2019kea}, we can estimate the total tau neutrino CC events based on the Eq.\ref{eq:ocp} with 5 years of run as 
\begin{equation}
N^{cc}_{\nu_\tau}\sim (3\times 10^4).
\end{equation}
Notice the annual expected tau neutrino yields is already comparable or even surpass the rates at the SHiP experiment at CERN~\cite{SHiP:2021nfo}. Thus our proposal can serve as a `brighter' neutrino factory for tau neutrinos. Here we would like to mention that with this abundance in neutrinos fluxes, it may possible that some of the new physics models, such as charged Higgs doublet~\cite{Branco:2011iw} or leptoquarks~\cite{deMedeirosVarzielas:2019lgb} maybe tested through new generation lepton collider~\cite{Qian:2022wxa} whose colliding beams are produced with the electron-positron collision.
 
\begin{figure}[h]
    \centering
    \includegraphics[scale=0.45]{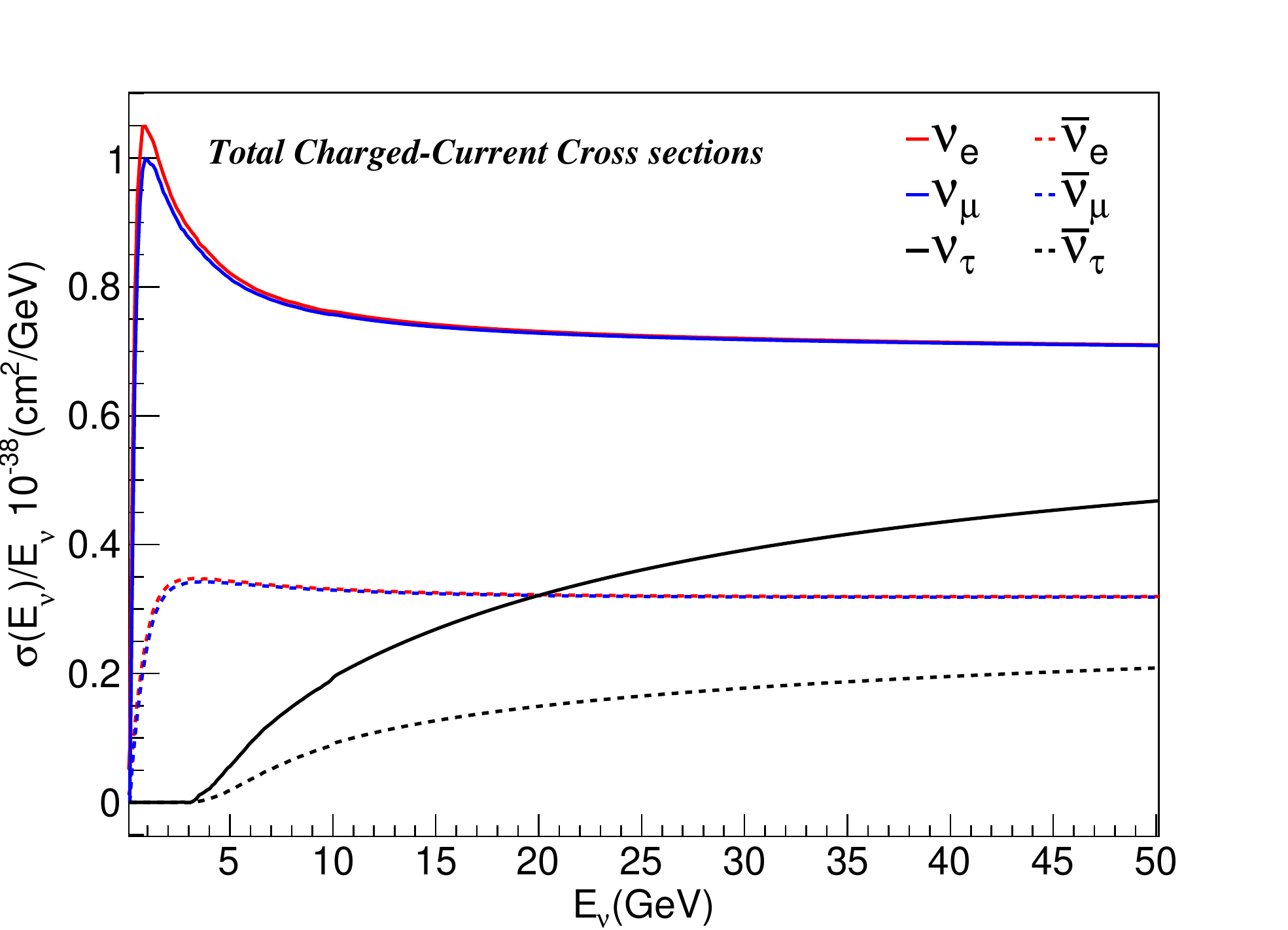}
    \caption{Neutrino and anti-neutrino cross sections with Argon Atom of mass number 40. The solid lines represent the cross sections of neutrino and nucleon interaction while the dashed lines are for anti-neutrino nucleon cross sections.}
    \label{fig:nsig}
\end{figure}

 Given the differences of the cross sections for neutrino and antineutrino, we simulate the interactions of neutrino with Argon atom with mass number 40 using \genie~\cite{Andreopoulos:2009rq, Andreopoulos:2015wxa, GENIE:2021zuu} event generator with the version 3.2.0 to obtain neutrino and anti-neutrino cross sections. Indeed, the cross sections for neutrino and antineutrino differ as the neutrino energy grows. Fig.~\ref{fig:nsig} shows the total Charged Current (CC) cross sections for three neutrino flavors and their anti-neutrino correspondents with respect to neutrino energy. Here the total cross sections should be understood as the sum of cross sections for available CC interactions of the neutrinos with nucleons of the Argon atom. The solid lines refer to neutrino-nucleon cross sections while the dashed lines represent cross sections of anti-neutrino nucleon interactions. The simulated results for cross sections are consistent with those of the DUNE's experimental configuration~\cite{DUNE:2021cuw} except for the $\nu_{\tau}$ cross sections because DUNE experiment mainly focuses on $\nu_{\mu}\rightarrow\nu_e$ oscillation channel. Clearly, the cross sections for $\nu_e$ and $\nu_{\mu}$ behave similarly, manifesting their cross sections are almost constant as the neutrino energy increases. However, the $\nu_{\tau}$ cross sections manifest a clear dependency on its energy. Another important feature is commonly present for both three neutrinos: cross sections for antineutrino  are approximately half the size or even smaller than the neutrino cross sections.

As the oscillation route for neutrinos is significantly longer than other short base line experiments \rm{i.e.} \texttt{MOMENT} ~\cite{Tang:2019wsv}, the matter effects during oscillations should be taken into account. To keep consistency for the neutrino oscillation probabilities, we utilize the matte-effected oscillation probability formulae coded within \globes. 

Looking at the latest neutrino oscillation parameters from~\cite{PDG:2022pth}, we set the oscillation parameters as shown in Tab.~\ref{tab:params}. The first column are the parameters used for obtaining the oscillation probability and event rates. The second column is the their central values while the third column shows the range of these parameters if marginalized. 
\begin{table}[ht]
\centering
\begin{tabular}{|c|c|c|}
\hline
\bf{Parameter} & \bf{True Value} & \bf{Marginalization Range} \\
\hline
$\sin^2 \theta_{12}$ & 0.310 & Not Marginalized \\
\hline
$\sin^2 \theta_{13}$ & 0.0241 & [0.01, 0.03] \\
\hline
$\sin^2 \theta_{23}$ & 0.58 & [0.38,0.64] \\
\hline
$\dcp$ & $[0, \pi]$ & $[-\pi, \pi]$ \\
\hline
$\Delta m^2_{21}$ (eV$^2$) & $7.39 \times 10^{-5}$ & Not Marginalized \\
\hline
$\Delta m^2_{31}$ (eV$^2$) & $2.449 \times 10^{-3}$ & Not Marginalized \\
\hline
\end{tabular}
\caption{The parameters used for neutrino oscillations: true values for data simulation in \globes are listed in the second column, while third column depicts the free and fixed variables for minimizing the $\chi^2$ function.}
\label{tab:params}
\end{table}

\section{Event distributions and CP violation sensitivity}\label{sec:results}
In this section, we show the simulated event distributions with the help of \globes. As stated earlier, we use muon and anti-muon beams produced from highly energetic positron collision with a high electron density target. Neutrinos then are produced from muon decay. We have four neutrino flavors that provide eight appearance channel. Here we are only interested in appearance channel because disappearance channel do not provide any sensitivity on $\dcp$. The spectrum and flavor content of neutrino beam are completely characterized by the muon decay. Therefore, with the knowledge of muon energy, 

we can control and obtain corresponding event spectrum. Luckily, we have here both muon and its CP-conjugated beams so that we can run both in the neutrino and anti-neutrino mode simultaneously. The symmetric operation of both beams leads to the cancellation or drastic reduction of many errors. Tab.~\ref{tab:expconfig} lists characteristic parameters for our simulation. Using available information from~\cite{Antonelli:2015nla, MuonCollider:2022nsa, MuonCollider:2022xlm}, we can estimate the total numbers of actively decaying muons that can be stored for neutrino production is up to $10^{20}\sim10^{21}$ order of magnitude. It is worth to mention that the total events for anti-neutrino production is visibly smaller than that of the neutrino at the detector side because of the differences between their cross sections. Thus we have to set a higher muon numbers for anti-neutrino mode or run longer time for anti-neutrino mode.  Based on the estimation method applied in \globes, the number of events in the i-th energy for transition from flavor $\alpha\rightarrow\beta$ are given by 
\begin{equation}
    N_i = \frac{N}{L^2}\int_0^{E_{max}} dE \int_{E_r^{min}}^{E_r^{max}}dE_r\phi(E)\sigma_{\nu_{\beta}} R(E, E_r)P_{\alpha\beta}(E)\epsilon(E_r),
    \label{eq:events}
\end{equation}
where $N$ is normalization factor proportional to run time and nuclear numbers in the target detector, $L$ is the length of the base line, $\phi(E)$ is the neutrino flux, $\nu_{\beta}$ is the neutrino interaction cross section, $\epsilon(E_r)$ and $R(E, E_r)$ are the efficiency and  the energy resolution function of the detector. The quantities $E$ and $E_r$ are the true and reconstructed energies respectively. For the detector properties, we utilize the energy resolutions and the detector efficiency provided by DUNE collaboration published on their \globes simulation~\cite{DUNE:2021cuw} in the case of electron and muon neutrino appearance rates. For tau neutrino event rates, we use a built-in energy resolution function based on a Gaussian distribution given as
\begin{equation}
    R^{\alpha}(E, E^{r}) = \frac{1}{\sigma(E)\sqrt{2\pi}} e^{-\frac{(E-E^{r})^2}{\sigma^2(E)}},
    \label{eq:eres}
\end{equation}
where $\sigma(E)$ is the efficiency we need and is taken to be approximately 80\% in reconstructed energy.
\begin{table}[ht]
\centering
\begin{tabular}{|c|c|}
\hline
\bf{Experimental Parameters} & \bf{Values}  \\
\hline
 Stored Muons & $1\times 10^{20}$  \\
\hline
$E_{\mu}$[GeV] & 22.5 GeV  \\
\hline
Run time & 5 years  \\
\hline
Matter density & 2.8 $g/cm^3$ \\
\hline
Base line length & 1300 Km \\
\hline
Target mass (Detector) & 40 Kt Liquid Argon \\
\hline
\end{tabular}
\caption{The experimental setup used in \globes for simulation.}
\label{tab:expconfig}
\end{table}

\begin{figure}[ht]
    \centering
    \includegraphics[scale=0.4]{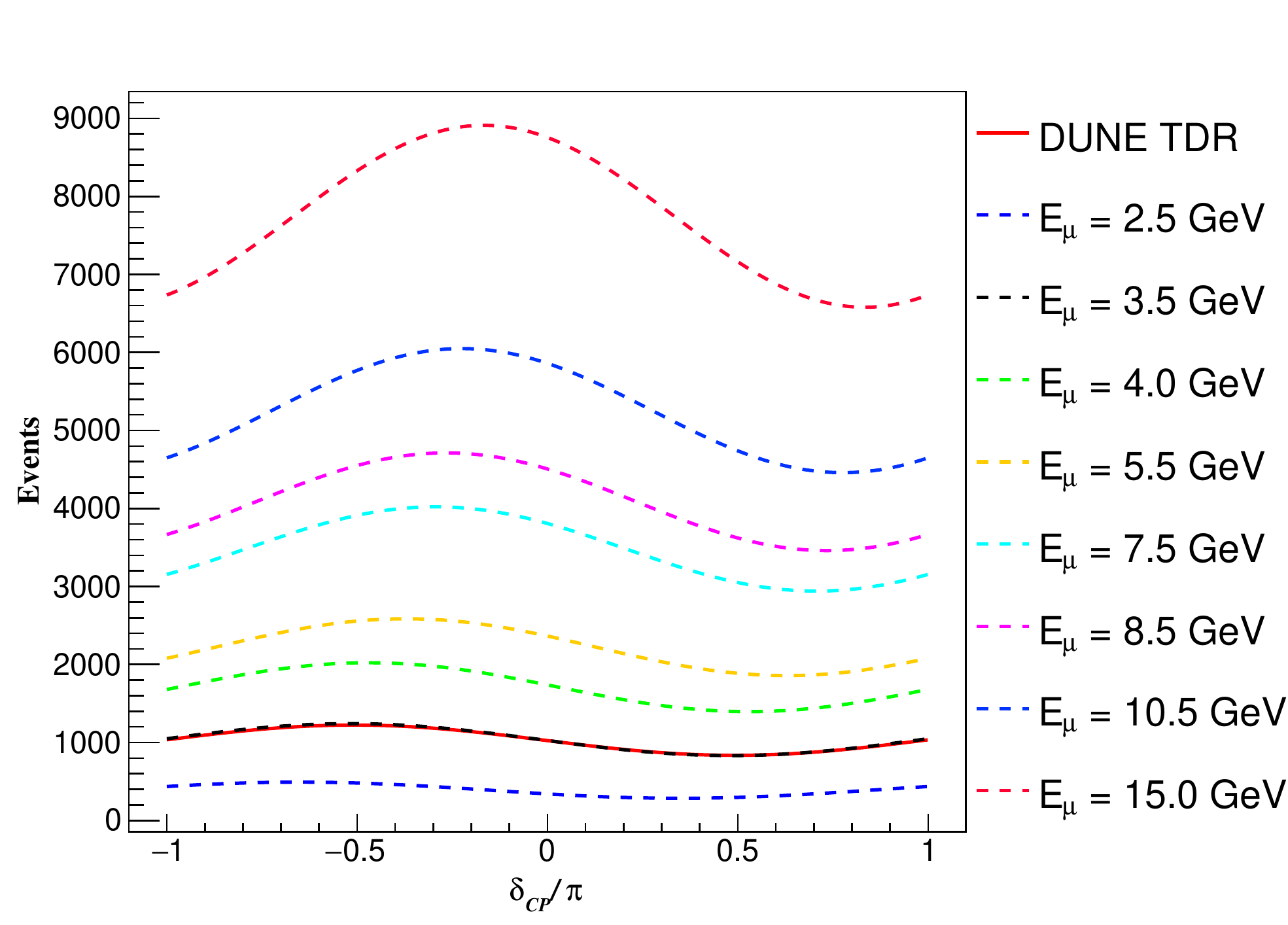}
    \includegraphics[scale=0.4]{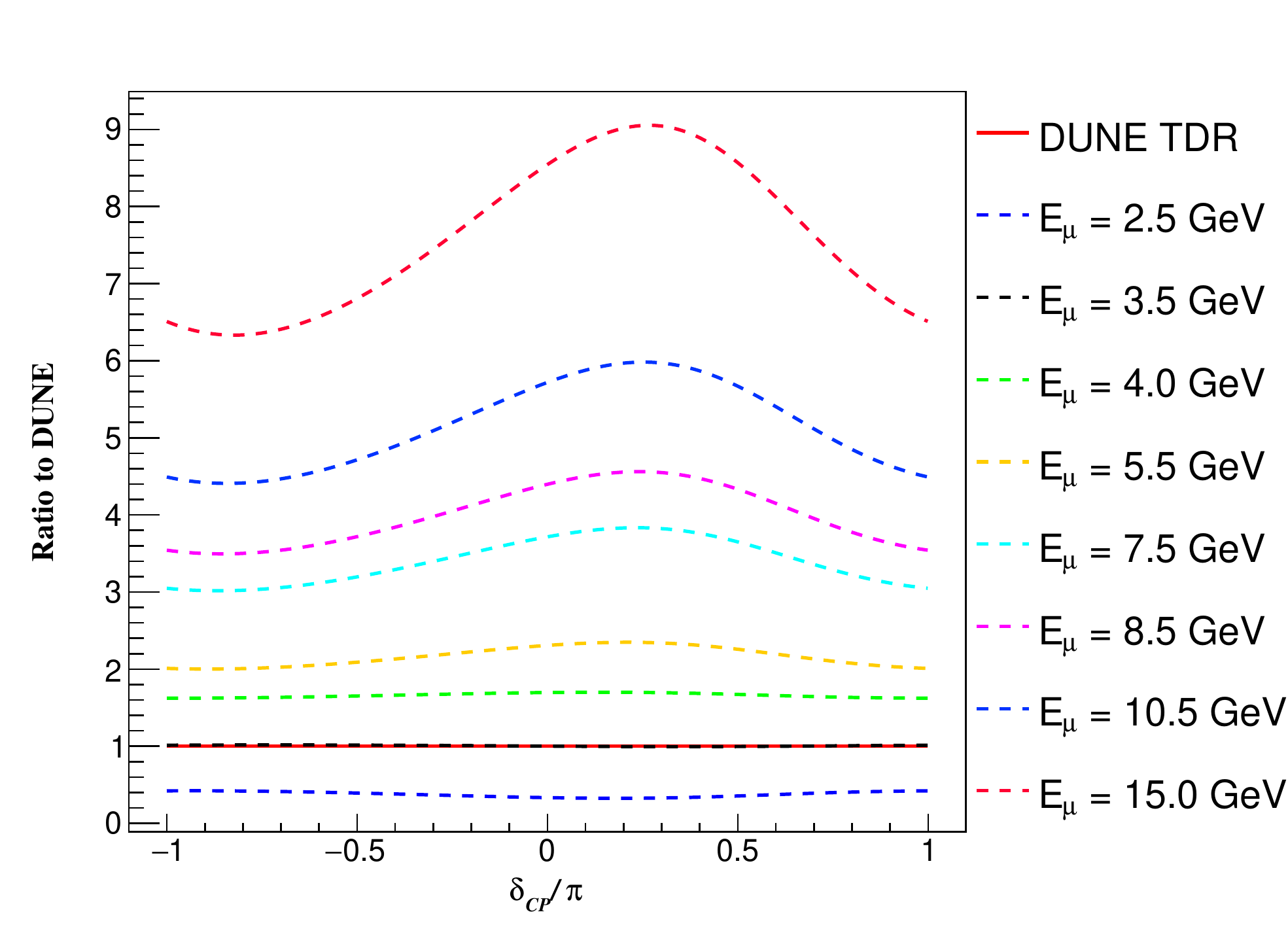}
    \caption{Event distribution as function of $\dcp$ (on the left) and the ratio of our results to that of the DUNE (on the right) for different muon energies are shown. Events are only taken for $\nu_e$ appearance from $\nu_{\mu}$ oscillation. The red solid line in both plots stand for the simulated results using DUNE's experimental design provided in~\cite{DUNE:2021cuw} while other dashed lines depict our results.}
    \label{fig:events_ratio}
\end{figure}

Fig.~\ref{fig:events_ratio} depicts event distribution as function of $\dcp$ for $\nu_e$ appearance from $\nu_{\mu}$ and the ratio of our results to that of the DUNE (solid red line) at different parent muon energy, ranging from 2.5 ${\rm GeV}$ to 15 ${\rm GeV}$. The red solid line corresponding to DUNE TDR is obtained by simulation using the DUNE's experimental configuration given in~\cite{DUNE:2021cuw}. We show this  because at the lower energy the oscillation patterns and thus the event spectrum for neutrinos in our case is identical to those of the DUNE experiment while they are visibly different toward higher energy. The event rates for electron neutrino obtained with our proposal is approximately seven to eight times larger than DUNE's $\nu_e$ event rates as shown on the right-hand plot. 

\begin{figure}[htbp]
    \centering
        \includegraphics[scale=0.44]{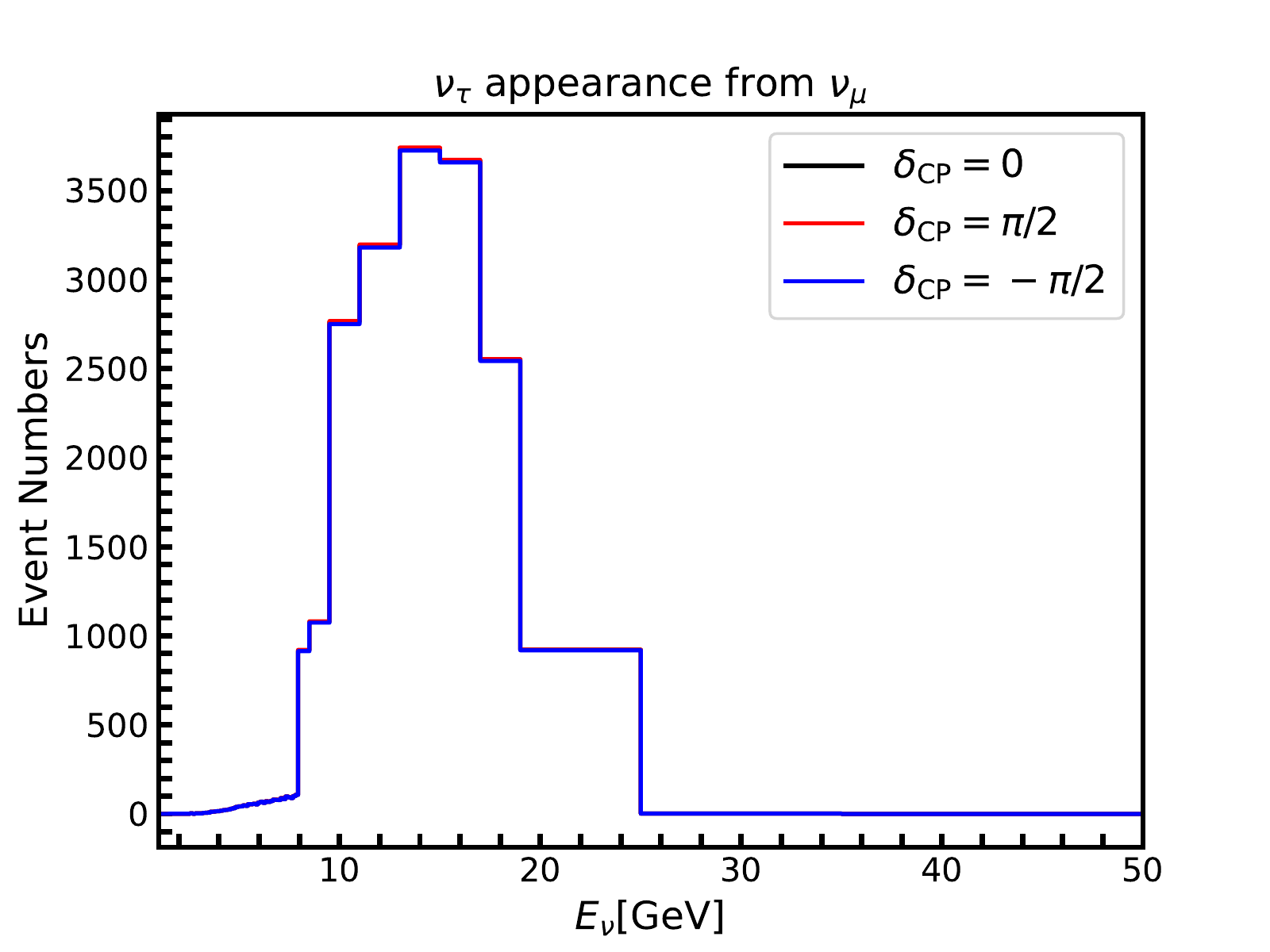}
        \includegraphics[scale=0.44]{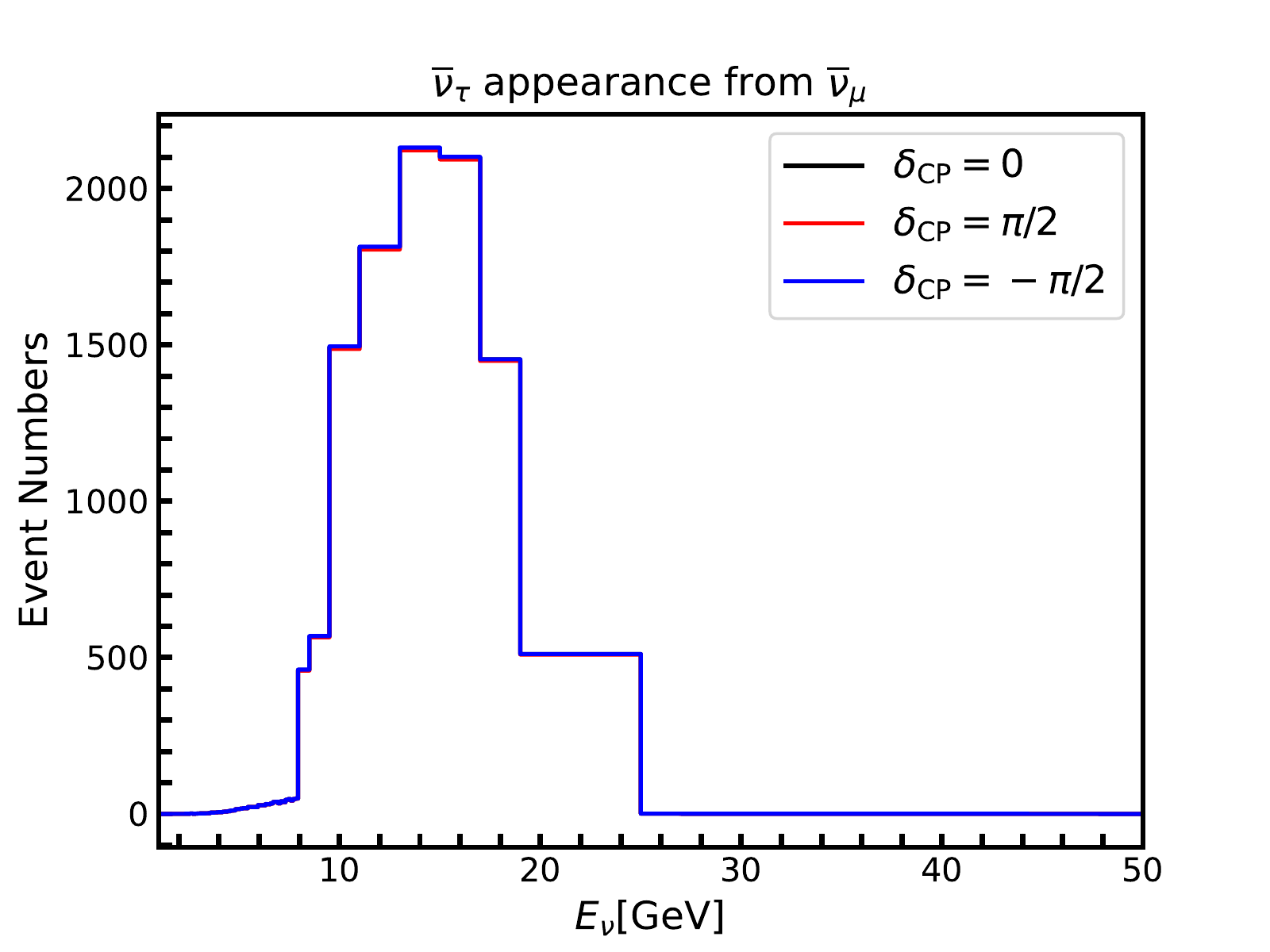}\\
        \includegraphics[scale=0.44]{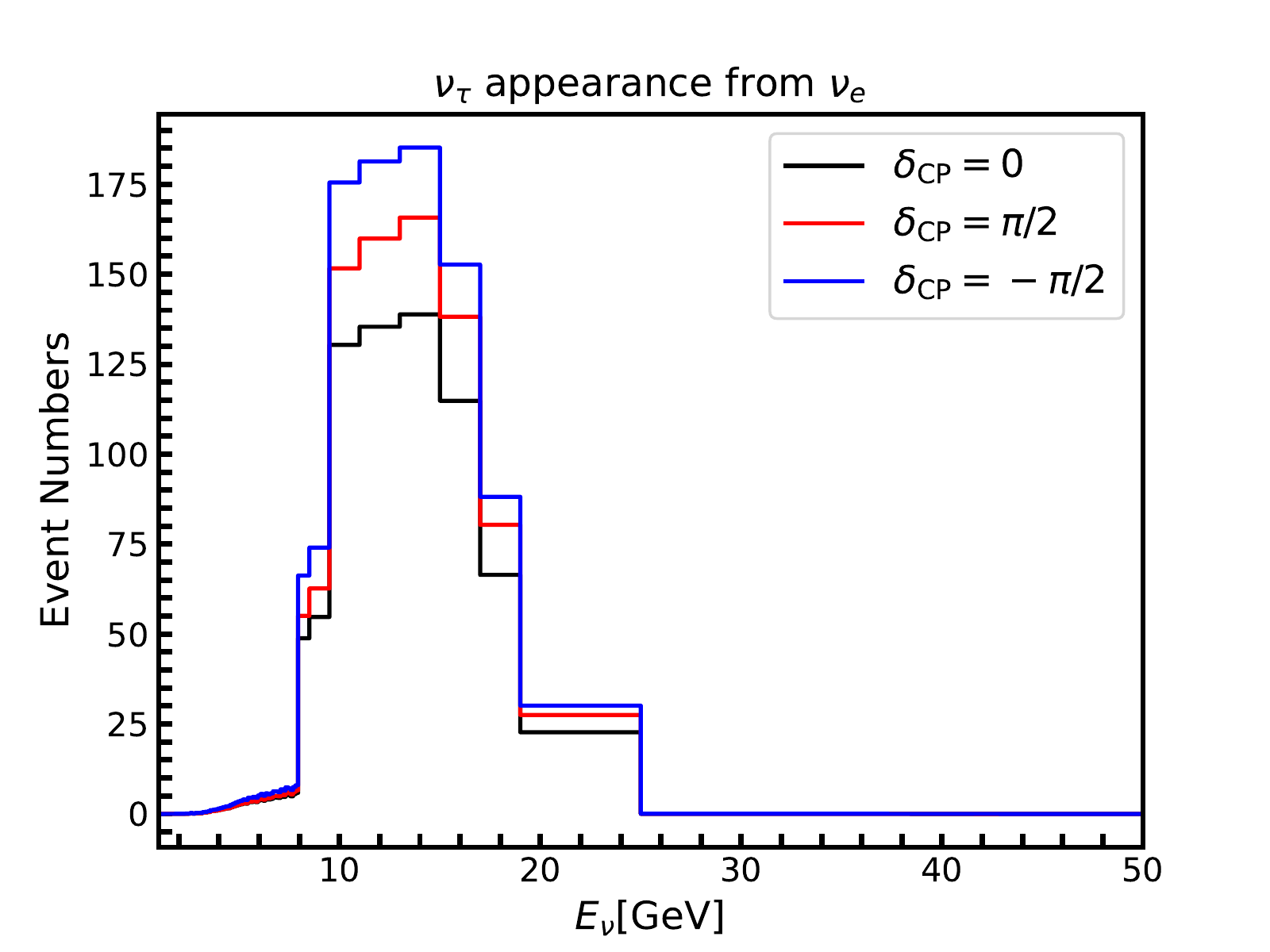}
        \includegraphics[scale=0.44]{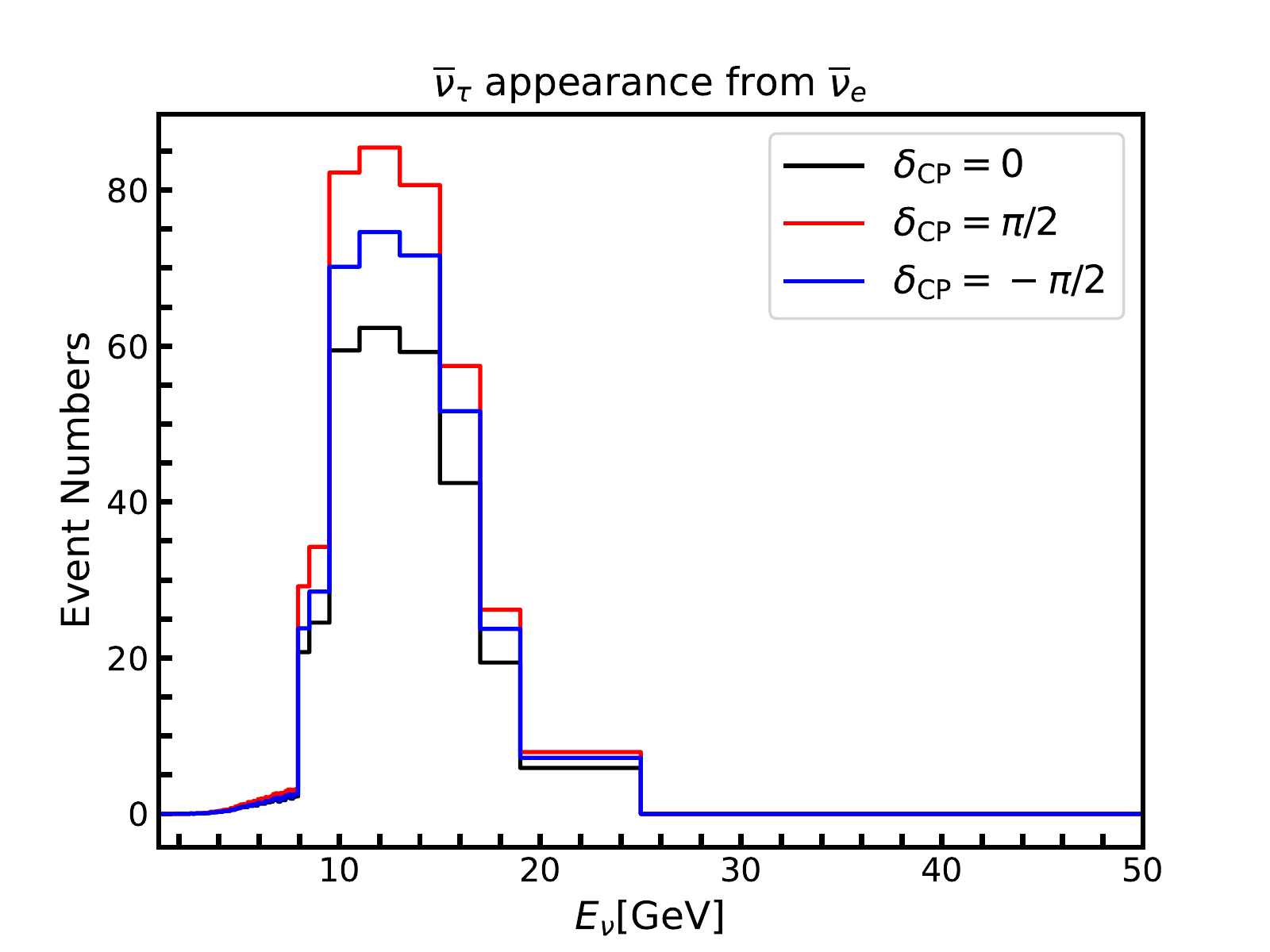}\\
        \includegraphics[scale=0.44]{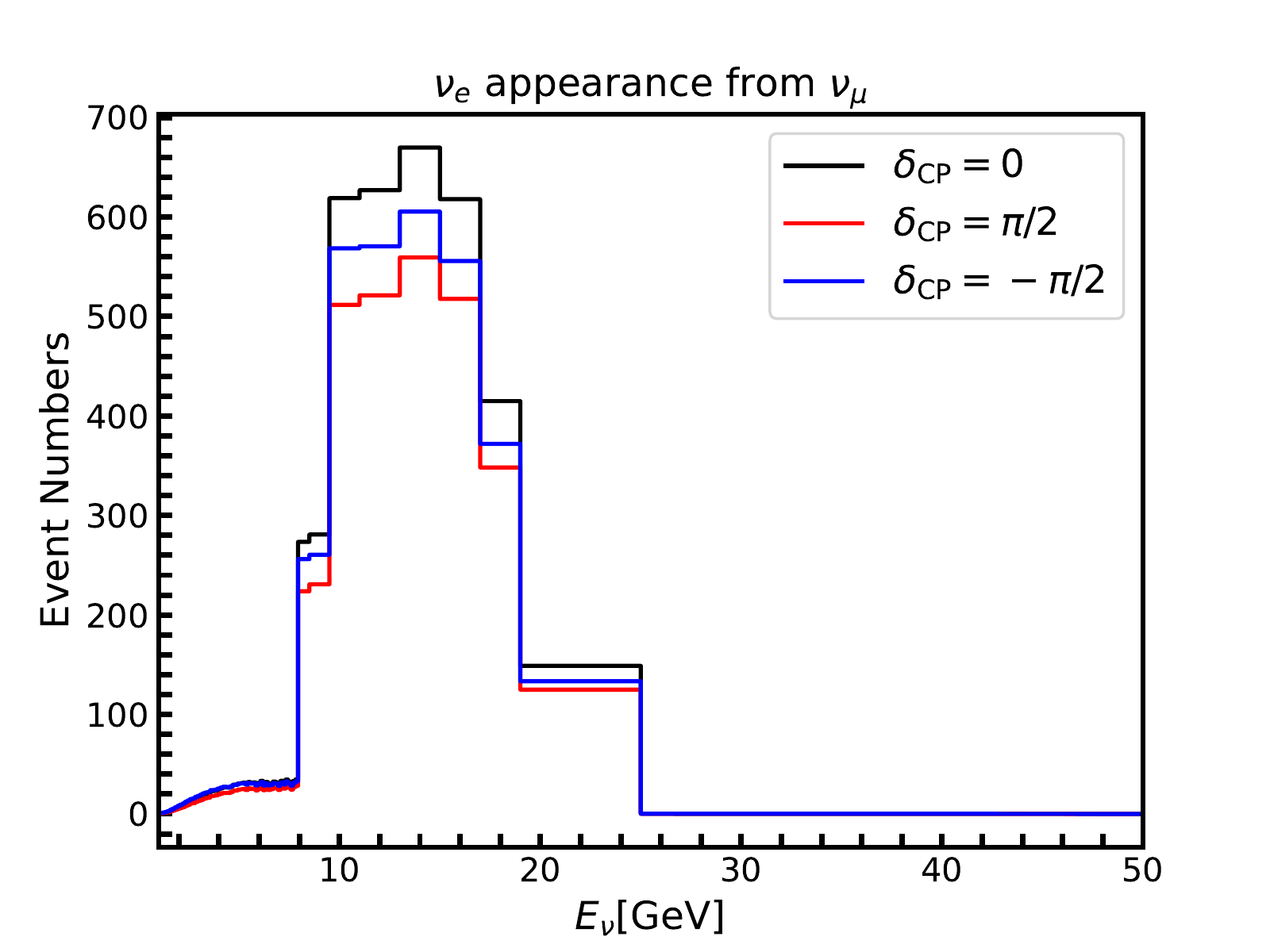}
        \includegraphics[scale=0.44]{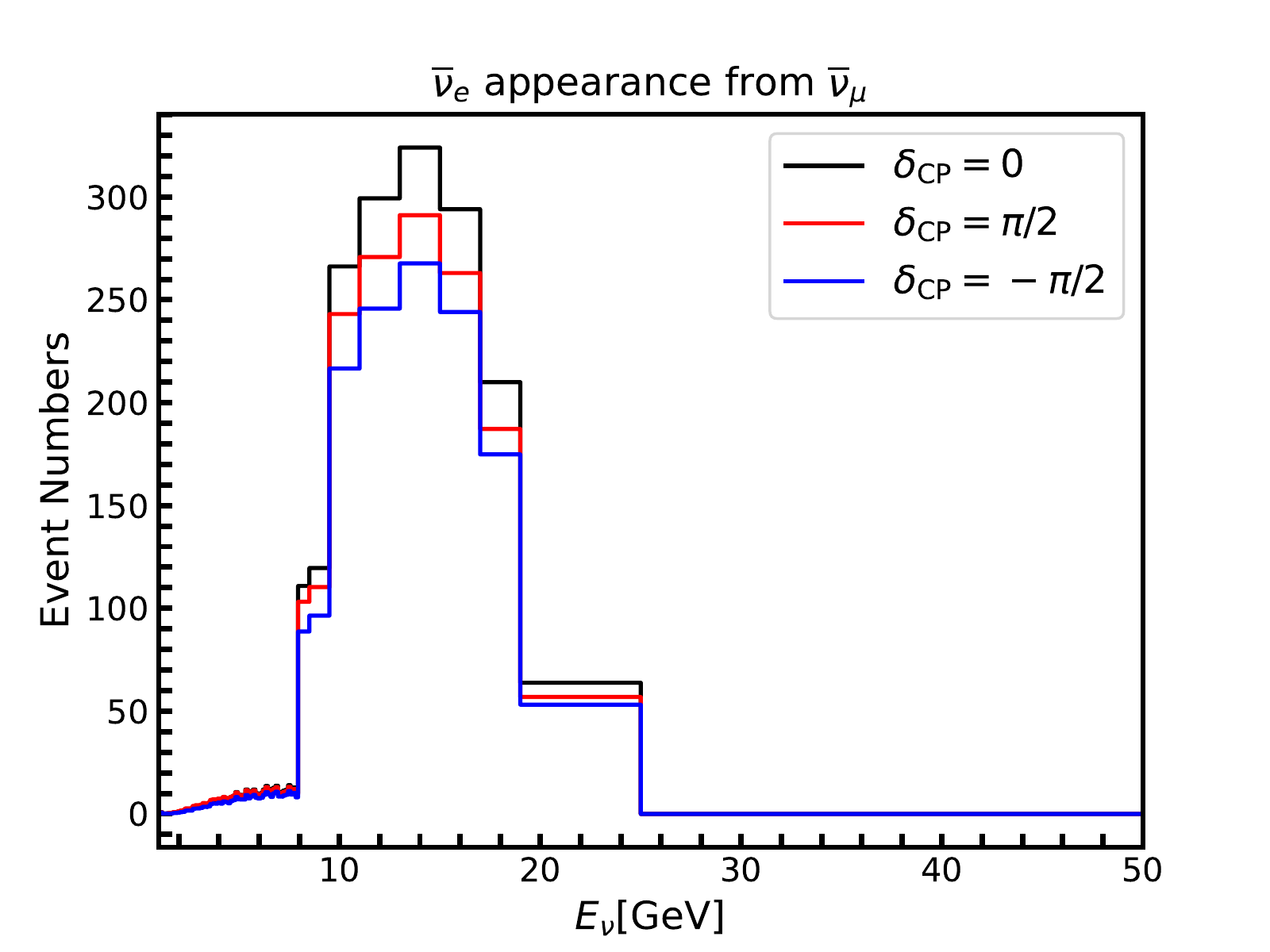}\\
        \includegraphics[scale=0.44]{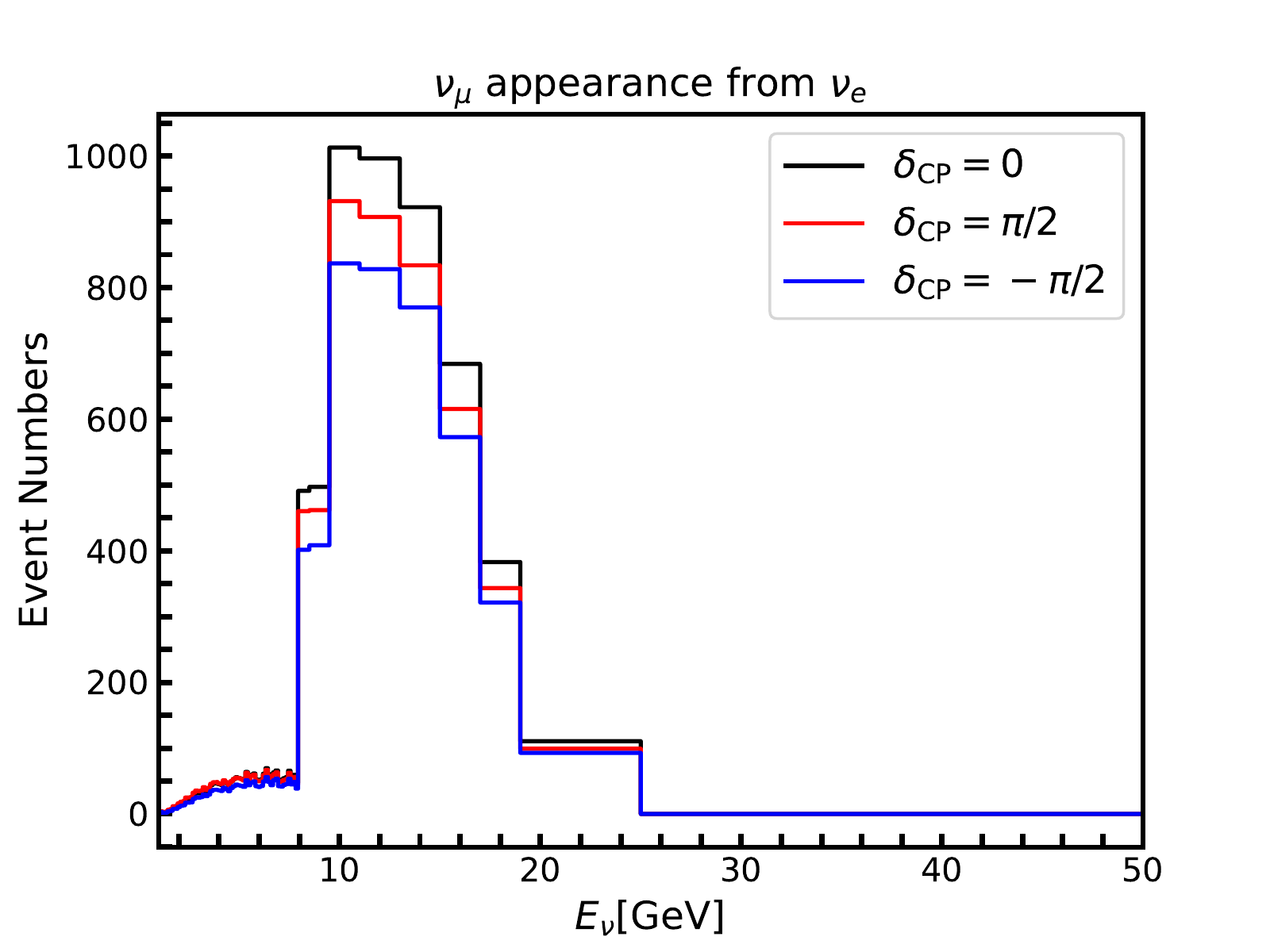}
        \includegraphics[scale=0.44]{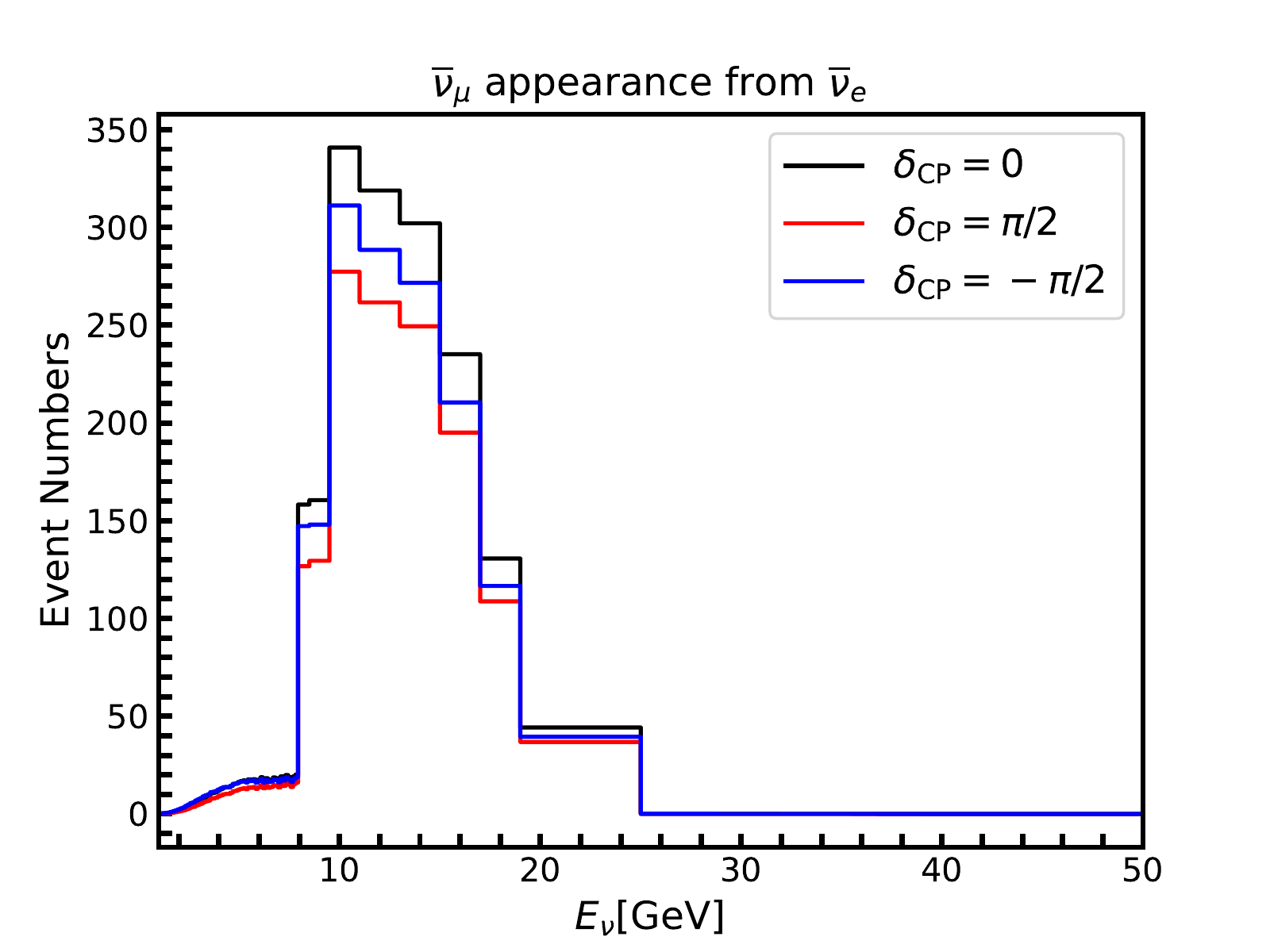}
    \caption{Neutrino (on the left) and antineutrino (on the right) event spectrum as function of reconstructed energy. Results are obtained with a 5-year of run simultaneously for neutrino and antineutrino mode with normal mass ordering is assumed for all the $\dcp$ values. The parameters and experiment characteristics are given on Tab.~\ref{tab:params} and Tab.~\ref{tab:expconfig}}
\label{fig:all_event}
\end{figure}

The event distributions as a function of reconstructed neutrino energy are shown in Fig~\ref{fig:all_event} for $\dcp = 0, \pi/2, -\pi/2$. The Maximal numbers of events for all channels have peaks approximately at 15 GeV.  From the shapes of event distributions from  $\nu_{\mu}\rightarrow \nu_{\tau}$ (on the top fist rows in Fig.~\ref{fig:all_event}), we can find that events of three different $\dcp$ values are not visibly different for $\nu_{\tau}$ appearance form $\nu_{\mu}$ oscillation, which makes the sensitivity from this channel weaker than other appearance channel. However, events from the other channels are visibly differentiable for $\dcp = 0, \pi/2, -\pi/2$, which is good for obtaining a high sensitivity for $\dcp$. As shown clearly, the event spectra for neutrino appearance are almost double of the number of anti-neutrino events since the cross section of neutrino interaction is significantly larger than the anti-neutrino cross sections. Here we like to point out that event distribution for $\dcp = 0$ is smaller or higher than the event distributions for $\dcp = \pi/2$ and $\dcp = -\pi/2$. This is different from the low-energy neutrino oscillation experiments,\eg, T2K, NOvA etc, whose event distributions are in  increasing (neutrino) or decreasing (antineutrino) order for $\dcp$ values. This is because of the different oscillation patterns at the high energy shown in the Fig.~\ref{fig:vacosc}.

$\chi^2$ analysis is performed by comparing the simulated true event rates from the present best fit~\cite{Esteban:2020cvm} with the events generated by the test values which is to be excluded. During the sensitivity calculation, only the solar parameters are allowed to vary while other parameter are kept fixed and constrained by Gaussian priors with $1 \sigma$ standard error. The prior functions are defined as 
\begin{equation}
    \chi^2_{\text{prior}} = \lp\frac{p - p_0}{\sigma_{p}} \rp^2,
\end{equation}
where $p$ is the oscillation parameter, $p_0$ is the central value of the prior measured by present experiment with absolute input error $\sigma_p$. In our analysis, we apply an uncertainty of $2.5\%$ for matter density. We construct the below $\chi^2$ function
\begin{equation}
    \chi^2 = \frac{\lp N(\dcp^{true}) - \overline{N}(\dcp^{true}) \rp^2}{(N + \overline{N})(\dcp = 0,\pi)}.
\end{equation}
Here, $N$ and $\overline{N}$  represent event rates for neutrino and anti-neutrino appearance.

\begin{figure}[ht]
    \centering
    \includegraphics[scale=0.4]{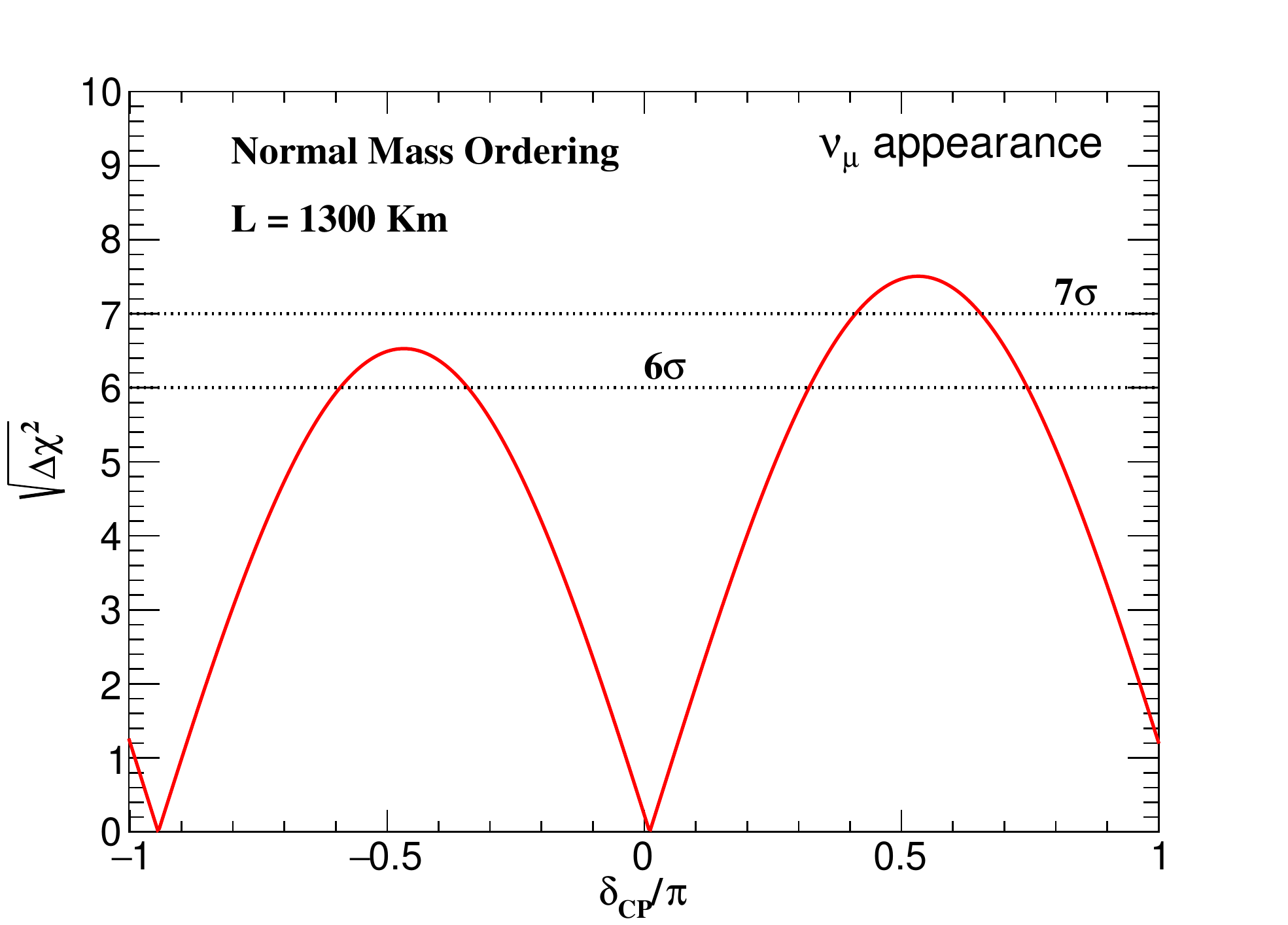}
    \includegraphics[scale=0.4]{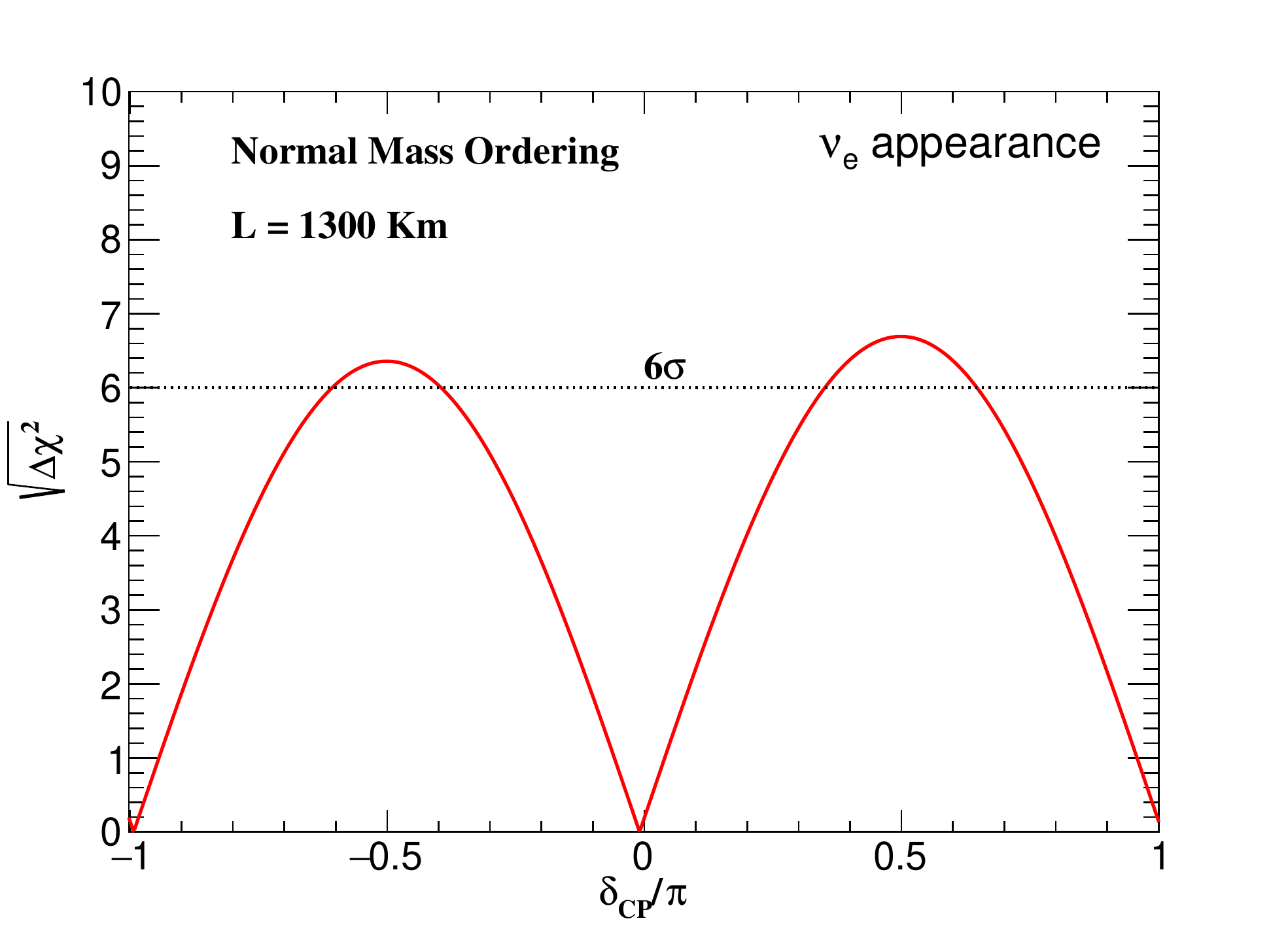}
    \includegraphics[scale=0.4]{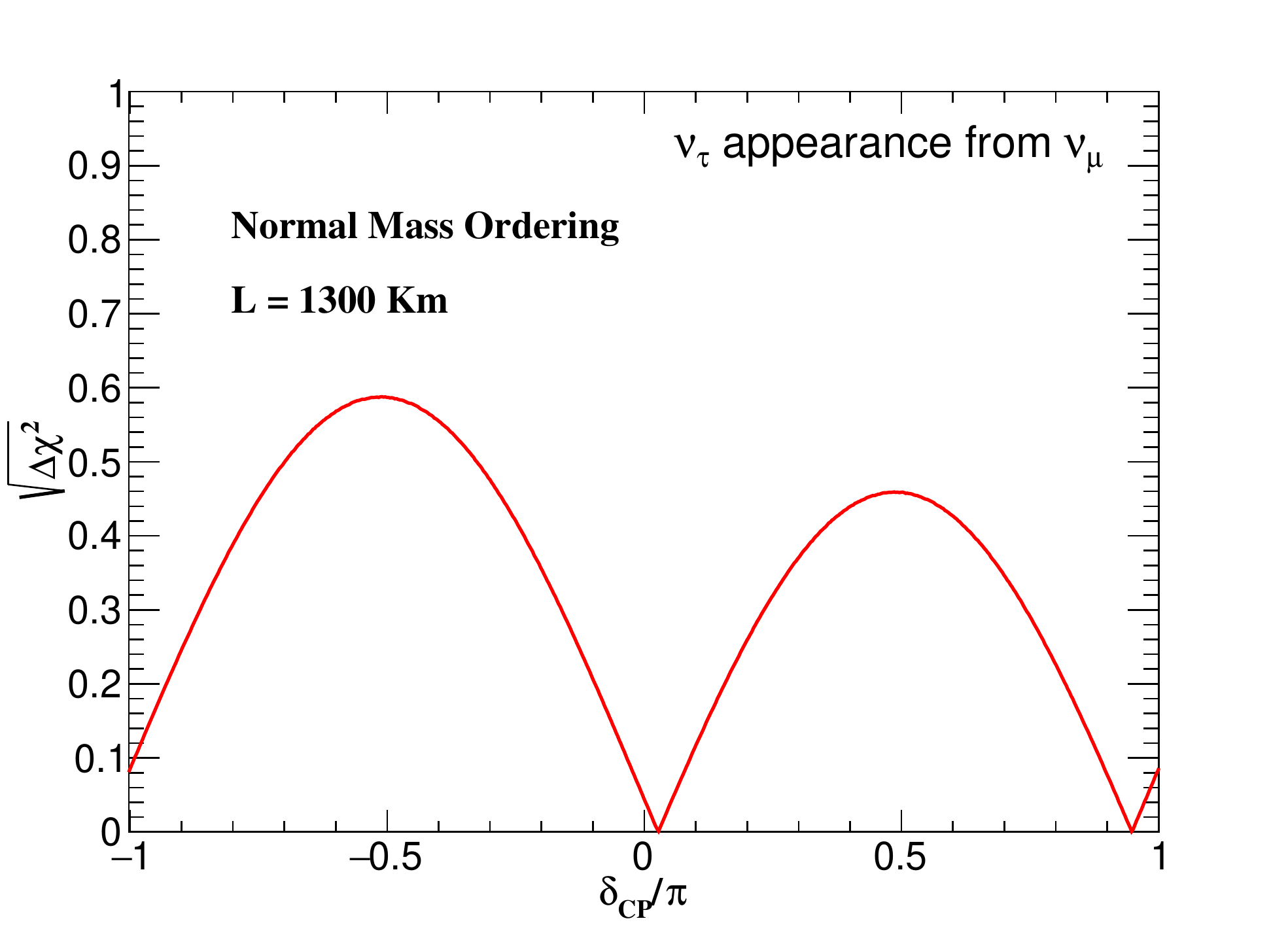}
    \includegraphics[scale=0.4]{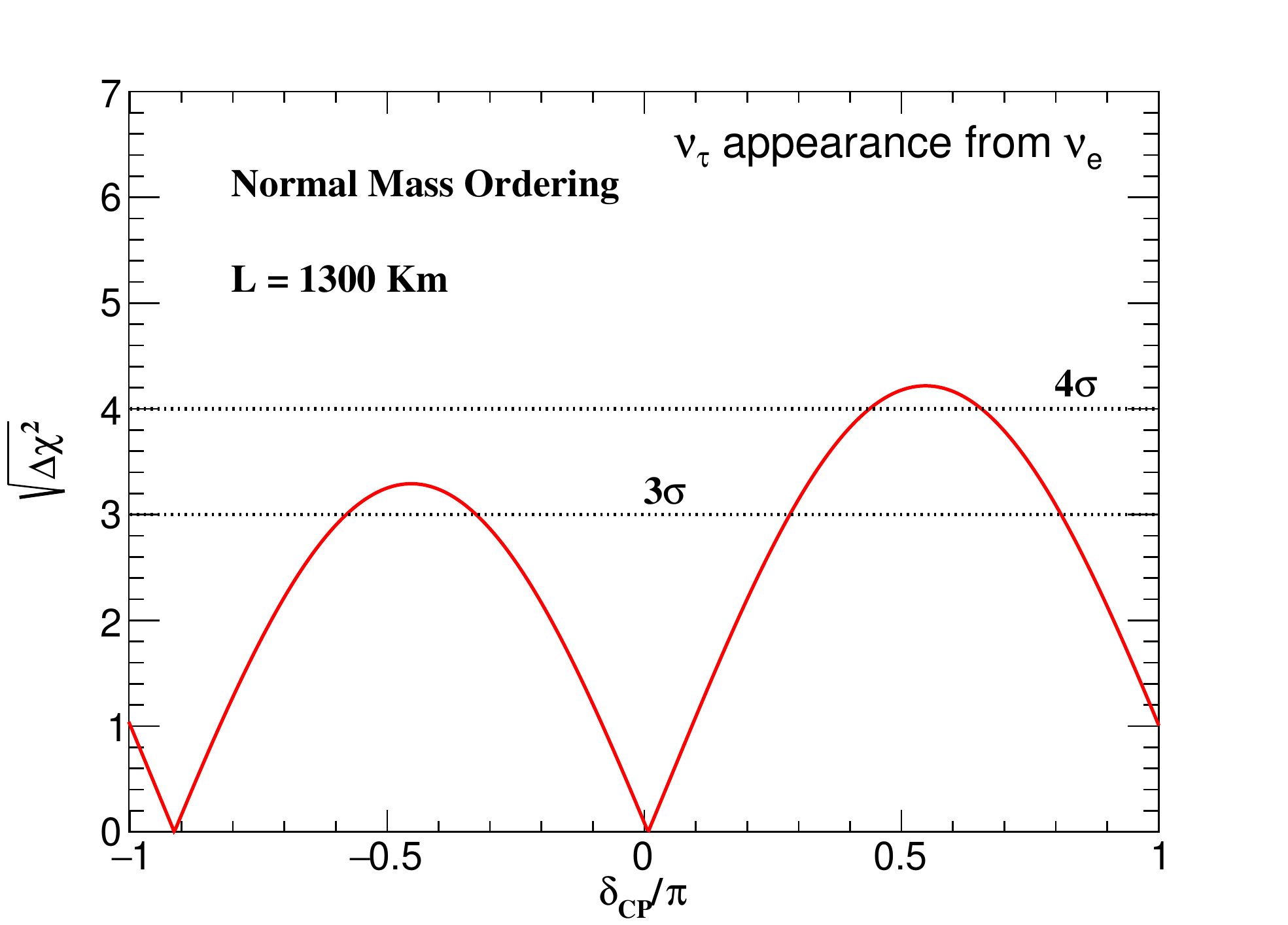}
    \caption{The significance for determination of CP violation as a function of the value of $\dcp$ for a simultaneous run of five years, assuming normal ordering.}
    \label{fig:cpsense}
\end{figure}

\begin{figure}[h]
    \centering
    \includegraphics[scale=0.45]{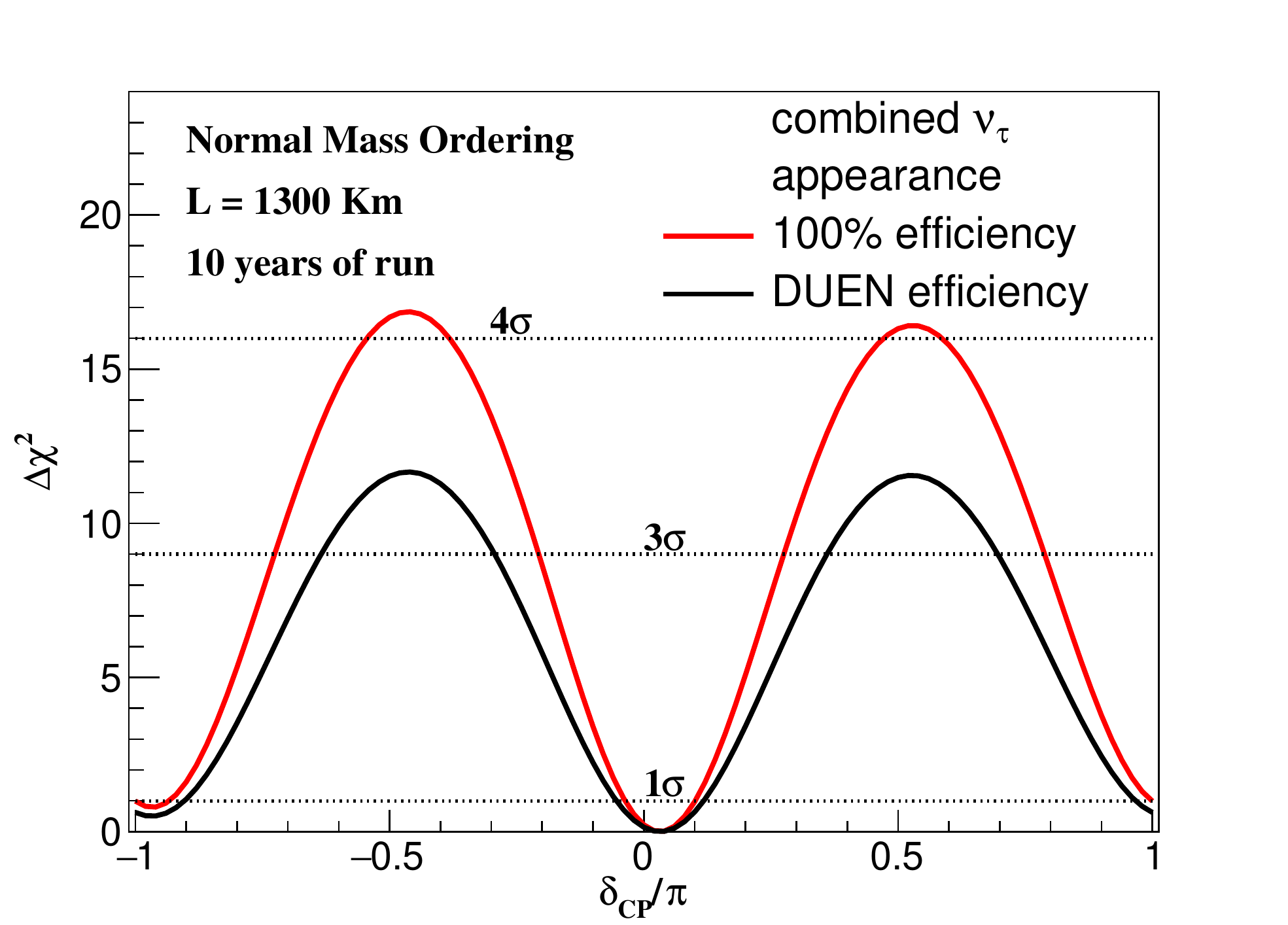}
    \caption{$\dcp$ sensitivity obtained with the fit from total $\nu_{\tau}$ appearance channel. Corresponding events are obtained with a 10-years of run assuming Normal mass Ordering. Here the red line represents obtained sensitivity with nearly 10\% efficiency in each energy bin while the red line represents sensitivity obtained using DUNE-provided energy efficiency that comes with their experimental configuration file.}
    \label{fig:total_tau}
\end{figure}

We will now evaluate the sensitivities on neutrino CP violation, taking $\delta_{\textrm{CP}}=\pm \pi/2$ as benchmark parameters.
\begin{itemize} 
\item Firstly, if the far detector has the capability of distinguishing electron and muon neutrino from one another, then both the $\nu_{\mu}$ and $\nu_e$ appearance channel can provide a high CP sensitivity. The sensitivity results are displayed in Fig.~\ref{fig:cpsense}.
The left-panel of the figure depicts estimated sensitivity for $\nu_{\mu}$ appearance channel while the figure on the right-hand side display estimated sensitivity for $\nu_e$ appearance channel. It is clear that $\nu_{\mu}$ appearance channel has the potential to discover  CP violation up to $7\sigma$ sensitivity at $\dcp = \pi/2$ while there is also sensitivity for $\dcp=-\pi $ and $\pi$ with more than $1\sigma$. On the other hand, the  $\nu_e$ appearance channel can also provide discovery of the maximum CP violation with more than 6$\sigma$ significance. 

\item Secondly, one of the advantages of our proposal is that we can also gain sensitivity in the case of detecting  $\nu_{\tau}$-related events. It is well-known fact that observing tau neutrino is extremely difficult. Luckily,  DUNE-type detectors can handle this problem. The bottom-left figure depicts the significance of $\nu_{\tau}$ appearance from $\nu_{\mu}$ oscillation. As seen, for five years of run, this sensitivity is very small because the event rate for $\dcp = 0, -\pi, \pi$ are not visibly different, see the first top row of Fig.~\ref{fig:all_event}. However, the bottom-right figure displays $\dcp$ sensitivity for the  $\nu_{\tau}$ appearance from $\nu_e$ oscillation channel. The corresponding significance can rich up to $4\sigma$ although this channel is not as good as the first two appearance channels for obtaining better sensitivity.  Notice that the CP dependence of $P(\nu_e \to \nu_\tau)$ and $P(\bar{\nu}_{\mu} \to \bar{\nu}_\tau)$ as shown in Eq.~\ref{eq:oscp} vary in the same direction. If we count on tau-related events in the far detector inclusively, this means $\nu_{\tau}$ signal can be further strengthened. The sensitivity then can be more than 4 standard deviations. Although only statistics are taken into account here, systematic uncertainty could be reduced efficiently due to the symmetric property of the proposed device. Furthermore, it is possible to exchange $\mu^+$ and $\mu^-$ flying routes, thus further reducing possible bias or systematic.

\item Finally, if we analyze total $\nu_{\tau}$ event spectrum from $\mu^-$ and $\mu^+$ beams without differentiation of electrons or muons in the detector, we can obtain sensitivity for maximum CP violation more than 4$\sigma$ with a 10-years of run. Fig.~\ref{fig:total_tau} displays obtained sensitivity for CP phase. 
\end{itemize}

\section{Potential for sterile neutrino search}\label{sec:nues}

The existence of the fourth neutrino or sterile neutrino is another major problem whose discovery may solve bunch of BSM mysteries. It is commonly known that the $Z$ gauge boson can decay into a pair of neutrino and antineutrino. Thus the measurements of the $Z$ boson decay width helps to determine the active neutrino numbers to be 3~\cite{ALEPH:2005ab}. However, there is possibility that the sterile neutrinos are allowed too given that they are the singlets of the SM gauge group and do not interact directly with SM gauge bosons. Generally, only one sterile neutrino $\nu_s$ is considered that consists mainly of the heavy mass eigenstate $\nu_4$ while the SM active neutrinos are mainly composed of the light neutrino mass eigenstates, $\nu_1, \nu_2, \nu_3$. Sterile neutrinos appear naturally in many extensions of SM. Meanwhile, there are several experimental hints of their existence. As is well known, the GALEX and SAGE solar neutrino Gallium experiments reported that only 88$\pm5\%$ $\nu_e$ events of the expected number were observed~\cite{Abdurashitov:2005tb,Kaether:2010ag}. The deficit of $\nu_e$ events observed in these  experiments can be explained by electron neutrino to sterile neutrino oscillation at short baseline.
The explanation of the LSND and MiniBooNE~\cite {LSND:1996ubh,MiniBooNE:2007uho,MiniBooNE:2022emn} experimental results could also indicate the possible existence of sterile neutrino. 
There is one more advantage of our proposal when it comes to search for the sterile neutrinos. The rich flux of both the muon and electron-type neutrinos produced after muon decay increases the possibility of observing oscillations related to sterile neutrino. We can examine two oscillation modes simultaneously: $\nu_e\to\nu_e$ and  $\nu_{\mu}\to\nu_e$, while DUEN and T2K mainly focus on electron neutrino appearance, $\nu_{\mu}\to \nu_e$.

The probability of disappearance and appearance for a neutrino flavor $\alpha$, taking into only the large mass difference account, can be approximated as
\begin{align}
    \label{eq:taust}
    P(\nu_{\alpha}\to \nu_{\alpha}) &\approx 1-4|U_{\alpha4}|^2(1-|U_{\alpha4}|^2)\sin^2\left( \frac{\Delta m_{41}^2 L}{4E_{\nu}} \right)\\
    P(\nu_{\alpha}\to \nu_{\beta}) &\approx
    4 |U_{\alpha 4}|^2|U_{\beta 4}|^2\sin^2\left(\frac{\Delta m_{41}^2L}{4E_{\nu}}\right)
\end{align}
With the fourth neutrino, the PMNS matrix is a $4\times4$ matrix that contains six mixing angles and three CP phases. Determination of these parameters would be a huge work that requires large number of neutrino oscillation experiments. According to some research work dealing with long-base line neutrino oscillations in the presence of the sterile neutrino~\cite{Reyimuaji:2019wbn}, our proposal not only enables us to further confirm the results of the LSND and MiniBooNE experiment but also help us to determine the aforementioned parameters, especially $\Delta m_{41}^2$, active and sterile neutrino mixing angles, as well as the additional CP phases. The detailed study for examining sensitivities of the parameters characterizing sterile neutrino will be presented in our future work.

\section{Conclusion and future outlook}\label{sec:summary}

In this work, we propose a new idea to exploit collimated muon beams (we take positron on target as an example method to produce those beams, but this can also be based on high energy muon beams from the proton on target method) which generate symmetric neutrino and antineutrino sources: $\mu^+\rightarrow e^+\,\bar{\nu}_{\mu}\, \nu_{e}$ and $\mu^-\rightarrow e^-\, \nu_{\mu} \,\bar{\nu}_{e}$. Interfacing with long baseline neutrino detectors such as DUNE or HyperK detectors, this experiment can be useful to measure tau neutrino properties, but importantly, to probe neutrino CP phase, by measuring $\nu_e, \nu_{\mu}$ and $\nu_{\tau}$ appearance, and differences between neutrino and antineutrino rates. By simulation using \globes software, the CP violation sensitivities with the appearance of all the three flavor neutrinos have been explored. Technically, there are several significant benefits leading to large neutrino flux and high sensitivity on CP phase. Firstly, the collimated and manipulable muon beams may lead to a larger acceptance of neutrino sources in the far detector side; Secondly, symmetric production of $\mu^+$ and $\mu^-$ beams also lead to symmetric neutrino and antineutrino production, which makes this proposal ideally good for measuring neutrino CP violation. More importantly, $\bar{\nu}_{e,\mu}\rightarrow\bar{\nu}_\tau$ and $\nu_{e,\mu}\rightarrow \nu_\tau$, and, $\bar{\nu}_{e}\rightarrow\bar{\nu}_\mu$ and $\nu_{e}\rightarrow \nu_\mu$ oscillation signals can be collected simultaneously without needs for separate specific runs for neutrinos or antineutrinos. So our experiment can be very time-saving . It is also possible to exchange $\mu^+$ and $\mu^-$ flying routes, thus further reducing possible bias or systematic.  The CP violation discovery in our approach is quite significant with focusing on the first two flavor neutrino appearance, more than 6$\sigma$ or for the total appearance that can rich to 7$\sigma$ within five years of run. But this requires to distinguish between electron and muon neutrinos, which can be done although difficult.

Moreover, The tension between T2K~\cite{T2K:2019bcf,Walsh:2022pqg} and  NOvA~\cite{NOvA:2021nfi} results on neutrino CP measurement may appear again, which makes an independent probe indispensable, for example, through tau neutrino appearance or electron to muon neutrino oscillations. The proposal here should also be useful to detect new CP phases in case of the presence of a sterile neutrino~\cite{deGouvea:2022kma}. Last but not least, our proposal exploits muon beam with looser requirement (e.g., lower intensity) compared with the needs toward a future muon collider, and thus can serve as a realistic intermediate step. 

In this draft, we mainly provide preliminary estimations (either qualitatively or based on \globes ) of the feasibility study. A more detailed study is surely necessary to follow up. On the other hand, there exist also rich potential to be further explored with such a proposal that connects energy and neutrino frontiers. Especially, one can imagine a post-DUNE (or in parallel to DUNE as the probe channels are indeed orthogonal and thus complementary) experiment with neutrinos from an intense muon source located at the Fermilab site. This connection between energy and neutrino frontiers can also serve as a precursor for future high-energy muon colliders. Notice that a muon collider requires a 1--2 orders of magnitude more intense beam as compared with the number ($dN_\mu/dt \sim 10^{12}$/sec ) listed above as our benchmark. Thus with the development of a more intensive muon beam targeting future muon colliders, it surely will improve further the neutrino potential of the current proposal.

\appendix
\begin{acknowledgments}
This work is supported in part by the National Natural Science Foundation of China under Grants No. 12150005, No. 12075004, and No. 12061141002, by MOST under grant No. 2018YFA0403900. The authors would like to thank Joachim Kopp, Haixing Lin and Jian Tang for useful discussions.
\end{acknowledgments}

\bibliographystyle{ieeetr}
\bibliography{paper_ref}

\end{document}